\pdfoutput=1

\documentclass[11pt,twoside,a4paper,cmspaper,final,collab]{cms-tdr}
\def\svnVersion{367291}\def\cmsCernNoTag{CERN-EP/2016-111}\def\cmsCernDate{\today}\def\cmsMessage{Published in the Journal of High Energy Physics as \href{http://dx.doi.org/10.1007/JHEP08(2016)122}{\doi{10.1007/JHEP08(2016)122.}}}

\begin{document}\cmsNoteHeader{SUS-15-007}

\hyphenation{had-ron-i-za-tion}
\hyphenation{cal-or-i-me-ter}
\hyphenation{de-vices}
\RCS$Revision: 367288 $
\RCS$HeadURL: svn+ssh://svn.cern.ch/reps/tdr2/papers/SUS-15-007/trunk/SUS-15-007.tex $
\RCS$Id: SUS-15-007.tex 367288 2016-09-10 20:28:23Z manuelf $

\newlength\cmsFigWidth
\ifthenelse{\boolean{cms@external}}{\setlength\cmsFigWidth{0.85\columnwidth}}{\setlength\cmsFigWidth{0.4\textwidth}}
\ifthenelse{\boolean{cms@external}}{\providecommand{\cmsLeft}{top\xspace}}{\providecommand{\cmsLeft}{left\xspace}}
\ifthenelse{\boolean{cms@external}}{\providecommand{\cmsRight}{bottom\xspace}}{\providecommand{\cmsRight}{right\xspace}}

\newcommand{\largr}{large-\ensuremath{R}\xspace}
\newcommand{\smalr}{small-\ensuremath{R}\xspace}
\newcommand{\cls}{\ensuremath{\mathrm{CL}_{\mathrm{S}}}\xspace}
\newcommand\zjets{{{\cPZ}+jets}\xspace}
\newcommand\wjets{{{\PW}+jets}\xspace}
\newcommand\ttjets{{\ensuremath{\ttbar}+jets}\xspace}
\newcommand{\MJ}{\ensuremath{M_J}\xspace}
\newcommand{\njets}{\ensuremath{N_{\text{jets}}}\xspace}
\newcommand{\nb}{\ensuremath{N_{\text{b}}}\xspace}
\newcommand{\mGlu}{\ensuremath{m_{\PSg}}\xspace}
\newcommand{\mLSP}{\ensuremath{m_{\PSGczDo}}\xspace}
\newcommand{\mStop}{\ensuremath{m_{\PSQt_1}}\xspace}
\newcommand{\mTop}{\ensuremath{m_{\PQt}}\xspace}
\newcommand{\mt}{\ensuremath{m_{\mathrm{T}}}\xspace}
\newcommand{\ghmdot}{\ensuremath{\,}}
\newcommand{\x}{\ensuremath{\phantom{0}}}
\cmsNoteHeader{SUS-15-007}
\title{Search for supersymmetry in pp collisions at $\sqrt{s} = 13\TeV$ in the single-lepton final state using the sum of masses of large-radius jets}

\date{\today}

\abstract{
Results are reported from a search for supersymmetric particles in proton-proton collisions
in the final state with a single, high transverse momentum lepton; multiple jets, including at least one b-tagged jet;
and large missing transverse momentum. The data
sample corresponds to an integrated luminosity of 2.3\fbinv at $\sqrt{s} = 13\TeV$,
recorded by the CMS experiment at the LHC. The search focuses on processes
leading to high jet multiplicities, such as gluino pair production with
${\PSg}\to{\ttbar}\,{\PSGczDo}$. The quantity $M_J$, defined as the sum of the masses
of the large-radius jets in the event, is used in conjunction with other kinematic variables to provide discrimination
between signal and background and as a key part of the background estimation method. The observed event yields in the
signal regions in data are consistent with those expected for standard model backgrounds, estimated from
control regions in data. Exclusion limits are obtained for a simplified model corresponding to gluino pair production
with three-body decays into top quarks and neutralinos. Gluinos with a mass below 1600\GeV are excluded
at a 95\% confidence level for scenarios with low $\PSGczDo$ mass, and neutralinos with a mass below 800\GeV are excluded for
a gluino mass of about 1300\GeV. For models with two-body gluino decays producing on-shell top squarks, the excluded region
is only weakly sensitive to the top squark mass.
}
\hypersetup{%
pdfauthor={CMS Collaboration},%
pdftitle={Search for supersymmetry in pp collisions at sqrt(s) = 13 TeV in the single-lepton final state using the sum of masses of large-radius jets},%
pdfsubject={CMS},%
pdfkeywords={CMS, physics, supersymmetry}}

\maketitle

\section{Introduction}
\label{sec:intro}

Supersymmetry
(SUSY)~\cite{Ramond:1971gb,Golfand:1971iw,Neveu:1971rx,Volkov:1972jx,Wess:1973kz,Wess:1974tw,Fayet:1974pd,Nilles:1983ge}
is an extension of the standard model (SM) of particle physics that is motivated by several considerations,
including the gauge hierarchy
problem~\cite{tHooft:1979bh,Witten:1981nf,Dine:1981za,Dimopoulos:1981au,Dimopoulos:1981zb,Kaul:1981hi}, the
existence of astrophysical dark matter~\cite{Zwicky:1933gu,Rubin:1970zza,Agashe:2014kda}, and the possibility of
gauge coupling constant unification at high energy~\cite{PhysRevD.24.1681,Sakai1981,IBANEZ1981439,EINHORN1982475,PhysRevD.25.3092}.
In SUSY models, each SM particle has a corresponding
supersymmetric partner (or partners) whose spin differs by one-half, such that fermions are mapped to bosons
and vice versa. Gauge quantum numbers are preserved by this symmetry, and to preserve degrees of freedom, a SM
spin-1/2 Dirac particle, such as the top quark, has two spin-0 partners, the top squarks. The SUSY partner of
the (spin-1) gluon, the massless mediator of the strong interactions in the SM, is the spin-1/2 gluino.
In $R$-parity--conserving models~\cite{Farrar:1978xj,Martin:1997ns}, SUSY particles are produced in
pairs, and the lightest supersymmetric particle (LSP) is stable.  If the LSP is the lightest neutralino
($\PSGczDo$), an electrically neutral mixture of the SUSY partners of the neutral electroweak gauge and Higgs bosons,
then it has weak interactions only and can in principle account for some or all of the dark matter.

The gauge hierarchy problem has become more urgent with the discovery of the Higgs
boson~\cite{Aad:2012tfa,Chatrchyan:2012ufa,Chatrchyan:2013lba,Khachatryan:2014jba,Aad:2014aba,Aad:2015zhl}.
Although the SM is conceptually complete, the Higgs boson mass, together with the electroweak scale, is
unstable against enormous corrections from loop processes, which pull the Higgs mass to the cutoff scale of
the theory, for example, the Planck scale.  This outcome can be avoided within the framework of the SM only
with extreme fine tuning of the bare Higgs mass parameter, a situation that is regarded as unnatural, although
not excluded. This problem suggests that additional symmetries and associated degrees of freedom may be
present that ameliorate these effects.  So-called natural SUSY
models~\cite{Dimopoulos:1995mi,Barbieri:2009ev,Papucci:2011wy,Feng:2013pwa}, in which sufficiently light SUSY
partners are present, are a major focus of current new physics searches at the CERN LHC. In natural models,
several of the SUSY partners are constrained to be light~\cite{Papucci:2011wy}: both top squarks, $\PSQt_L$
and $\PSQt_R$, which have the same electroweak couplings as the left- ($L$) and right- ($R$) handed top
quarks, respectively; the bottom squark with $L$-handed couplings ($\PSQb_L$); the gluino ($\PSg$); and the
Higgsinos ($\widetilde\Ph$).  While the gluino mass is not constrained by naturalness considerations as strongly
as that of the lighter top squark mass eigenstate, $\PSQt_1$, the cross section for gluino pair production is
substantially larger than that for top squark pair production, for a given mass. As a consequence, the two
types of searches can have comparable sensitivity to these models. Both types of searches are currently of
intense interest, and CMS and ATLAS data taken at $\sqrt{s}=8\TeV$ have provided significant
constraints~\cite{Craig:2013cxa} on natural SUSY scenarios.

This study uses the first LHC proton-proton collision data taken by the CMS experiment at $\sqrt{s} = 13\TeV$
to search for gluino pair production. Searches targeting this process in the single-lepton final state using 8\TeV data have been performed by both ATLAS~\cite{Aad:2014lra,Aad:2015mia} and CMS~\cite{Chatrchyan:2013iqa}.  For $\mGlu=
1.5$\TeV, somewhat above the highest gluino masses excluded at $\sqrt{s} = 8$\TeV, the cross section for gluino
pair production increases dramatically with center-of-mass energy, from about 0.4\unit{fb} at $\sqrt{s} = 8$\TeV to
about 14\unit{fb} at $\sqrt{s}=13$\TeV~\cite{Borschensky:2014cia}. In contrast, the cross section for the dominant
background, $\ttbar$ production, increases much more slowly, from about 248\unit{pb} at $\sqrt{s} =8$\TeV to 816\unit{pb}
at $\sqrt{s}=13$\TeV~\cite{ref:ttbarXSec}.  As a consequence, the sensitivity of this search can be
significantly extended with respect to searches performed at $\sqrt{s} = 8\TeV$, even though the 13\TeV data
sample has an integrated luminosity of only 2.3\fbinv, roughly one-tenth of that acquired at 8\TeV.

\begin{figure}[btp!]
\centering
\includegraphics[width=0.49\textwidth]{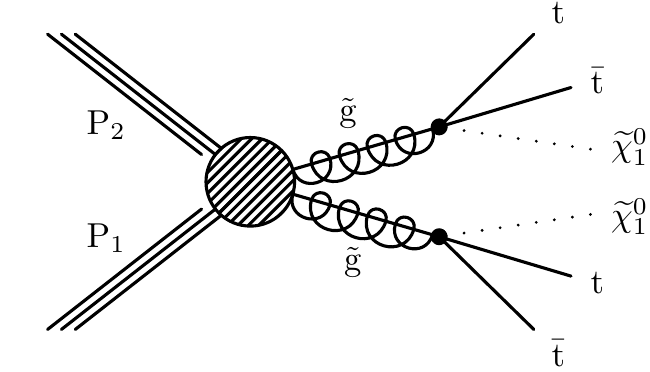}
\includegraphics[width=0.49\textwidth]{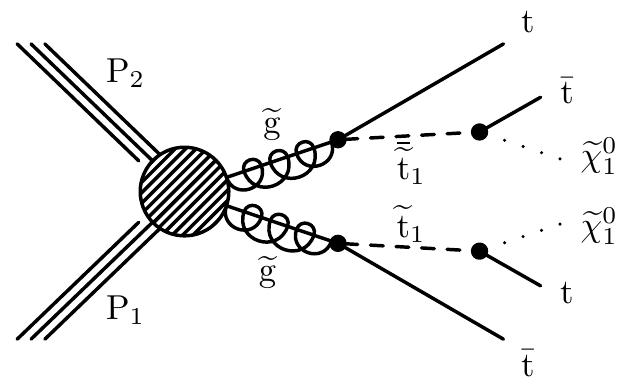}
\caption{Gluino pair production and decay for the simplified models T1tttt (left) and T5tttt (right). In T1tttt,
the gluino undergoes three-body decay $\PSg\to\ttbar\PSGczDo$ via a virtual intermediate top squark. In T5tttt, the
gluino decays via the sequential two-body process $\PSg\to\PSQt_1\PAQt$, $\PSQt_1\to\PQt\PSGczDo$.
Because gluinos are Majorana particles, each one can decay to $\PSQt_1\PAQt$ and to
the charge conjugate final state $\overline{\PSQt}_1\PQt$.
}
\label{fig:intro:nlep_t1tttt}
\end{figure}

The search targets gluino pair production with $\PSg\to\ttbar\PSGczDo$, which arises from
$\PSg\to\PSQt_1\PAQt$, where the top squark is produced either on or off mass shell.  The off-mass-shell
scenario is shown in Fig.~\ref{fig:intro:nlep_t1tttt} (left) and is often designated
T1tttt~\cite{Chatrchyan:2013sza} in simplified model scenarios~\cite{bib-sms-2,bib-sms-3,bib-sms-4}.  Results
are also obtained for scenarios with on-shell top squark masses. This scenario is shown in
Fig.~\ref{fig:intro:nlep_t1tttt} (right) and will be denoted by T5tttt.
(For this scenario, the small contribution from
the direct production of top squark pairs is also taken into account.)
Regardless of whether the top squark
is produced on or off mass shell, the final state is characterized by a large number of jets, four of which
are b jets from top quark decays.  Depending on the decay modes of the accompanying $\PW$ bosons, a range of
lepton multiplicities is possible; we focus here on the single-lepton final state, where the lepton is either
an electron or a muon.  Because the two neutralinos ($\PSGczDo$) are undetected, their production in SUSY events
typically gives rise to a large amount of missing (unobserved) momentum, whose value in the
direction transverse to the beam axis can be inferred from the momenta of the observed particles.
The missing transverse momentum, $\ptvecmiss$, is a key element of
searches for $R$-parity-conserving SUSY, and its magnitude is denoted by $\MET$.

A challenge in performing searches for SUSY particles is obtaining sufficient
sensitivity to the signal, while at the same time understanding the background
contribution from SM processes in a robust manner.  This analysis is designed
such that the background in the signal regions arises largely from a single
process, dilepton $\ttbar$ production, in which both $\PW$ bosons from
$\cPqt\to\cPqb\PW^+$ decay leptonically, but only one lepton satisfies the
criteria associated with identification, the minimum transverse momentum ($\pt$) requirement,
and isolation from other energy in the event. The search signature is characterized
not only by the presence of high-$\pt$ jets and b-tagged
jets, an isolated high-$\pt$ lepton, and large $\MET$, but also by additional kinematic
variables. Apart from resolution effects, the transverse mass of the lepton +
$\ptvecmiss$ system, \mt, is bounded above by $m_{\mathrm{W}}$
for events with a single leptonically decaying $\PW$, and this variable
is very effective in suppressing the otherwise dominant single-lepton
$\ttbar$ background. The quantity \MJ, the scalar sum of the masses of large-radius jets,
is used both to characterize the mass and energy scale of the event,
providing discrimination between signal and background, and as a key part of the
background estimation.  A property of $\MJ$ exploited in this analysis is
that, for the dominant background, this variable is nearly uncorrelated with \mt.
Because of the absence of correlation between $\MJ$ and $\mt$,
the background shape at high \mt, including the signal region, can be measured to
a very good approximation using a low-$\mt$ control sample.  The quantity $\MJ$ was
first discussed in phenomenological studies, for example, in
Refs.~\cite{Hook:2012fd,Cohen:2012yc,Hedri:2013pvl}.
Similar variables have been used by ATLAS for SUSY searches in all-hadronic final states using 8\TeV
data~\cite{Aad:2015lea,Aad:2013wta}. We have presented studies
of basic $\MJ$ properties and performance using early 13\TeV data~\cite{SUSYDPS}.

This paper is organized as follows. Section~\ref{sec:detector} gives a brief overview of the CMS detector.
Section~\ref{sec:samples} discusses the simulated event samples used in the
analysis. The event reconstruction is discussed in Section~\ref{sec:objects}, while
Section~\ref{sec:eventSelection} describes the trigger and event selection.
Section~\ref{sec:backgroundEstimation} presents the
methodology used to predict the SM background from the event yields in control regions in data.
The associated systematic uncertainties are also discussed.
The event yields observed in the signal regions are presented
in Section~\ref{sec:observationsInterpretation}. These yields are compared with background
predictions and used to obtain exclusion regions for the gluino pair production
models shown in Fig.~\ref{fig:intro:nlep_t1tttt}. Finally, Section~\ref{sec:summary}
presents a summary of the methodology and the results.

\section{Detector}
\label{sec:detector}

The central feature of the CMS detector is a superconducting solenoid of 6\unit{m} internal diameter,
providing a magnetic field of 3.8\unit{T}. Within the solenoid volume are the tracking and
calorimeter systems. The tracking system, composed of silicon-pixel and silicon-strip
detectors, measures charged particle trajectories within
the pseudorapidity range $\abs{\eta}<2.5$, where $\eta\equiv -\ln[\tan(\theta/2)]$ and $\theta$ is the
polar angle of the trajectory of the particle with respect to the counterclockwise proton beam direction.
A lead tungstate crystal electromagnetic calorimeter (ECAL), and a brass and scintillator hadron calorimeter (HCAL),
each composed of a barrel and two endcap sections, provide energy measurements up to $\abs{\eta}=3$.
Forward calorimeters extend the pseudorapidity coverage provided by the barrel and
endcap detectors up to $\abs{\eta}=5$. Muons are identified and measured within the range $\abs{\eta}<2.4$
by gas-ionization detectors embedded in the steel magnetic flux-return yoke outside the solenoid.
The detector is nearly hermetic, permitting the accurate measurement of $\ptvecmiss$.
A more detailed description of the CMS detector, together with a definition of the coordinate
system used and the relevant kinematic variables, is given in Ref.~\cite{Chatrchyan:2008zzk}.

\section{Simulated event samples}
\label{sec:samples}

The analysis makes use of several simulated event samples for modeling the SM background and signal processes.
While the background estimation in the analysis is performed largely from control samples in the data,
simulated event samples provide correction factors, typically near unity. The equivalent integrated luminosity of the
simulated event samples is at least six times that of the data, and at least 100 times that of the data in the
case of \ttbar and signal processes.

The production of \ttjets, \wjets, \zjets, and QCD multijet events is simulated with the Monte Carlo (MC)
generator \textsc{MadGraph5\_aMC@NLO}~2.2.2~\cite{Alwall:2014hca} in leading-order (LO) mode. Single top quark
events are modeled at next-to-leading order (NLO) with \textsc{MadGraph5\_aMC@NLO} for the $s$-channel and
\POWHEG v2~\cite{powheg-singletop-tchan,powheg-singletop-wt} for the $t$-channel and \PW-associated production.
Additional small backgrounds, such as \ttbar production in association with bosons, diboson processes, and
$\mathrm{t\bar{t}t\bar{t}}$ are similarly produced at NLO with either \textsc{MadGraph5\_aMC@NLO} or \POWHEG.
All events are generated using the NNPDF~3.0~\cite{Ball:2014uwa} set of parton distribution functions (PDF).
Parton showering and fragmentation are performed with the \PYTHIA~8.205~\cite{Sjostrand:2014zea} generator with
the underlying event model based on the CUETP8M1 tune detailed in Ref.~\cite{Khachatryan:2110213}. The
detector simulation is performed with \GEANTfour~\cite{Agostinelli:2002hh}. The cross sections used to scale
simulated event yields are based on the highest order calculation available. For \ttbar, in addition to using the
next-to-next-to-leading order + next-to-next-to-leading logarithmic cross section
calculation~\cite{ref:ttbarXSec}, the modeling of the event
kinematics is improved by reweighting the top quark $\pt$ spectrum to match the data~\cite{Khachatryan:2016641},
keeping the overall normalization fixed.

Signal events for the T1tttt and T5tttt simplified models are generated in a manner similar to that for the SM
backgrounds, with the
\textsc{MadGraph5\_aMC@NLO}~2.2.2 generator in LO mode using the NNPDF~3.0 PDF set and followed with
\PYTHIA~8.205 for showering and fragmentation.
The detector simulation is performed with the CMS fast simulation package~\cite{Abdullin:2011zz}
with scale factors applied to account for any differences with respect to the full simulation used for
backgrounds. Event samples are generated for a representative set of model scenarios by scanning over the
relevant mass ranges for the $\PSg$ and $\PSGczDo$, and the yields are normalized to the NLO + next-to-leading-logarithmic cross
section~\cite{Borschensky:2014cia,Beenakker:1996ch,Kulesza:2008jb,Kulesza:2009kq,Beenakker:2009ha}.

Throughout this paper, two T1tttt benchmark models are used to illustrate typical signal behavior.  The
T1tttt(1500,100) model, with masses $\mGlu=1500$\GeV and $\mLSP=100\GeV$, corresponds to a scenario with a
large mass splitting (referred to as non-compressed, or NC) between the gluino and the neutralino. This mass combination probes
the sensitivity of the analysis to a low cross section (14\unit{fb}) process that has a hard \MET spectrum, which
results in a relatively high signal efficiency. The T1tttt(1200,800) model, with masses $\mGlu=1200$\GeV and
$\mLSP=800\GeV$, corresponds to a scenario with a small mass splitting (referred to as compressed, or C) between the gluino and
the neutralino. Here the cross section is much higher (86\unit{fb}) because the gluino mass is lower than for
the T1tttt(1500,100) model, but the sensitivity suffers from a low signal efficiency due to the soft \MET spectrum.

Finally, to model the presence of additional proton-proton collisions from the same or adjacent beam crossing
as the primary hard-scattering process (``pileup'' interactions), the simulated events are overlaid with
multiple minimum bias events, which are also generated with the \PYTHIA~8.205
generator with the underlying event model based on the CUETP8M1 tune. The distribution of the number of overlaid
minimum bias events is broad and peaks in the range 10--15.

\section{Event reconstruction}
\label{sec:objects}

The reconstruction of physics objects in an event proceeds from the candidate
particles identified by the particle-flow (PF)
algorithm~\cite{cms-pas-pft-09-001,cms-pas-pft-10-001}, which uses information
from the tracker, calorimeters, and muon systems to identify the candidates as
charged or neutral hadrons, photons, electrons, or muons.  Charged particle tracks are
required to originate from the event primary vertex (PV), defined as
the reconstructed vertex, located within 24\unit{cm} (2\unit{cm}) of the
center of the detector in the direction along (perpendicular to) the beam
axis, that has the highest value of $\pt^2$ summed
over the associated charged particle tracks.

The charged PF candidates associated with the PV and the neutral PF candidates are
clustered into jets using the anti-$\kt$ algorithm~\cite{Cacciari:2008gp}
with distance parameter $R=0.4$, as implemented in the \textsc{fastjet} package~\cite{Cacciari:2011ma}.
The estimated pileup contribution to the jet $\pt$
from neutral PF candidates is removed with a correction based on the area
of the jet and the average energy density of the event~\cite{Cacciari:2007fd}.
The jet energy is calibrated using $\pt$- and $\eta$-dependent corrections; the resulting
calibrated jet is required to satisfy $\pt>30\GeV$ and $\abs{\eta}\leq2.4$.  Each
jet must also meet loose identification requirements~\cite{Chatrchyan:2011ds}
to suppress, for example, calorimeter noise. Finally, jets that have PF
constituents matched to an isolated lepton, as defined below, are removed from
the jet collection.

A subset of the jets are ``tagged'' as originating from \PQb quarks using the
combined secondary vertex (CSV) algorithm~\cite{Chatrchyan:2012jua,CMS:2016kkf}.
For the CSV medium working point chosen for
this analysis, the signal efficiency for $\PQb$ jets in the range $\pt=30$ to 50\GeV
is 60--67\% (51--57\%) in the barrel (endcap), increasing with \pt. Above $\pt\approx 150\GeV$
the $\PQb$ tagging efficiency decreases.
The probability to misidentify jets arising from $\PQc$ quarks
is 13--15\% (11--13\%) in the barrel (endcap), while the misidentification probability for light-flavor quarks
or gluons is 1--2\%.

Throughout this paper, quantities related to the number of jets ($\njets$) or to the
number of $\PQb$-tagged jets ($\nb$) are based only on small-$R$ jets, not on
the large-$R$ jets discussed below.

Electrons are reconstructed by associating a charged particle track with an ECAL
supercluster~\cite{Khachatryan:2015hwa}. The resulting candidate electrons are
required to have $\pt>20\GeV$ and $\abs{\eta}<2.5$, and to satisfy identification
criteria designed to remove light-parton jets, photon conversions, and
electrons from heavy flavor hadron decays.
Muons are reconstructed by associating tracks in the muon system with
those found in the silicon tracker~\cite{Chatrchyan:2012xi}. Muon candidates
are required to satisfy $\pt>20\GeV$ and $\abs{\eta}<2.4$.

To preferentially select leptons that originate in the decay of $\PW$ bosons,
leptons are required to be isolated from other PF candidates.
Isolation is quantified using an optimized version of the ``mini-isolation'' variable
originally suggested in Ref.~\cite{Rehermann:2010vq}, in which the transverse energy of
the particles within a cone in $\eta$-$\phi$ space surrounding the lepton momentum vector is computed
using a cone size that scales as $1/\pt^{\ell}$, where $\pt^{\ell}$ is the transverse
momentum of the lepton. In this analysis,
mini-isolation, $I_\text{mini}^\text{rel}$, is defined as the transverse energy of particles
in a cone of radius $R^{\text{mini-iso}}$ around the lepton,
divided by $\pt^{\ell}$.
The transverse energy is computed as the scalar sum of the \pt values of the charged hadrons from
the PV, neutral hadrons, and photons.
The neutral hadron and photon contributions to this sum are corrected for pileup.
The cone radius $R^{\text{mini-iso}}$ varies with the $\pt^{\ell}$ according to
\begin{equation}
  R^{\text{mini-iso}} =\begin{cases} 0.2, &
  \pt^{\ell}\leq 50\GeV\\ (10\GeV)/\pt^{\ell}, &
  \pt^{\ell}\in(50\GeV, 200\GeV)\\ 0.05, & \pt^{\ell}\geq
  200\GeV.
  \end{cases}\label{eqn:mini_iso_r}
\end{equation}
The $1/\pt^{\ell}$ dependence is motivated by considering a two-body decay
of a massive parent particle with mass $M$ and large $\pt$,
for which the angular separation of the daughter particles is roughly
$\Delta R_{\text{daughters}}\approx 2M/\pt.$
The \pt-dependent cone size reduces the rate of accidental overlaps between the
lepton and jets in high-multiplicity or highly Lorentz-boosted events, particularly
overlaps between \PQb jets and leptons originating from a boosted top
quark. The cone remains large enough to contain \PQb-hadron decay products for
non-prompt leptons across a range of $\pt^{\ell}$ values. Muons (electrons) must satisfy
$I_\text{mini}^\text{rel}<0.2$\,(0.1).
The combined efficiency for the electron
reconstruction and isolation requirements is about 50\% at a $\pt^{\ell}$ of 20\GeV,
increasing to 65\% at 50\GeV and reaching a plateau of 80\% above 200\GeV.
The combined reconstruction and isolation efficiencies for muons are
about 70\% at a $\pt^{\ell}$ of 20\GeV, increasing to 80\% at 50\GeV and reaching a plateau of 95\%
at 200\GeV.

We cluster $R=0.4$ (``\smalr'') jets and the isolated leptons into $R=1.2$
(``\largr'') jets using the anti-$\kt$ algorithm. The mass of the \largr jets
retains angular information about the clustered objects, as well as their \pt
and multiplicity. Clustering \smalr jets instead of PF candidates incorporates
the jet pileup corrections, thereby reducing the
dependence of the mass on pileup. The variable \MJ is defined as the sum of all
\largr jet masses:
\begin{equation}
  \MJ = \sum_{J_i = {\text{large-}R\text{ jets}}} m(J_i).
\end{equation}
The technique of clustering small-$R$ jets into large-$R$ jets has been used
previously by ATLAS in, for example, Ref.~\cite{Aad:2014bva}.
Leptons are included in the large-$R$ jets to include the full kinematics of the event, and
the choice $R=1.2$ optimizes the background rejection power of
\MJ while retaining signal efficiency. Larger distance parameters were found to offer no
significant additional discriminating power, while smaller parameters decrease the background rejection
up to a factor of two for models with small mass splittings between the gluino and neutralino.

\begin{figure}[tbp!]
  \centering
  \includegraphics[width=0.49\textwidth]{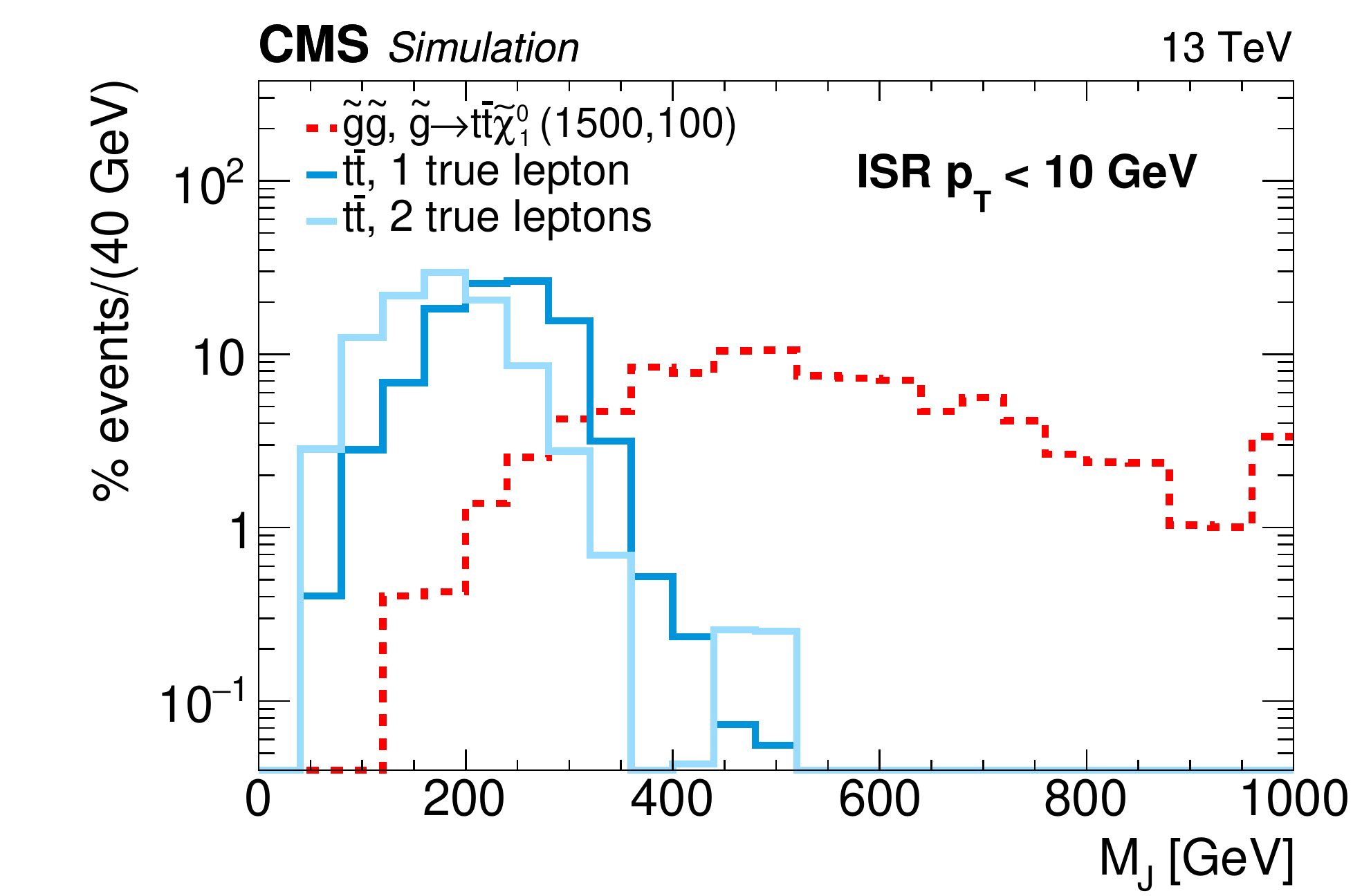}
  \includegraphics[width=0.49\textwidth]{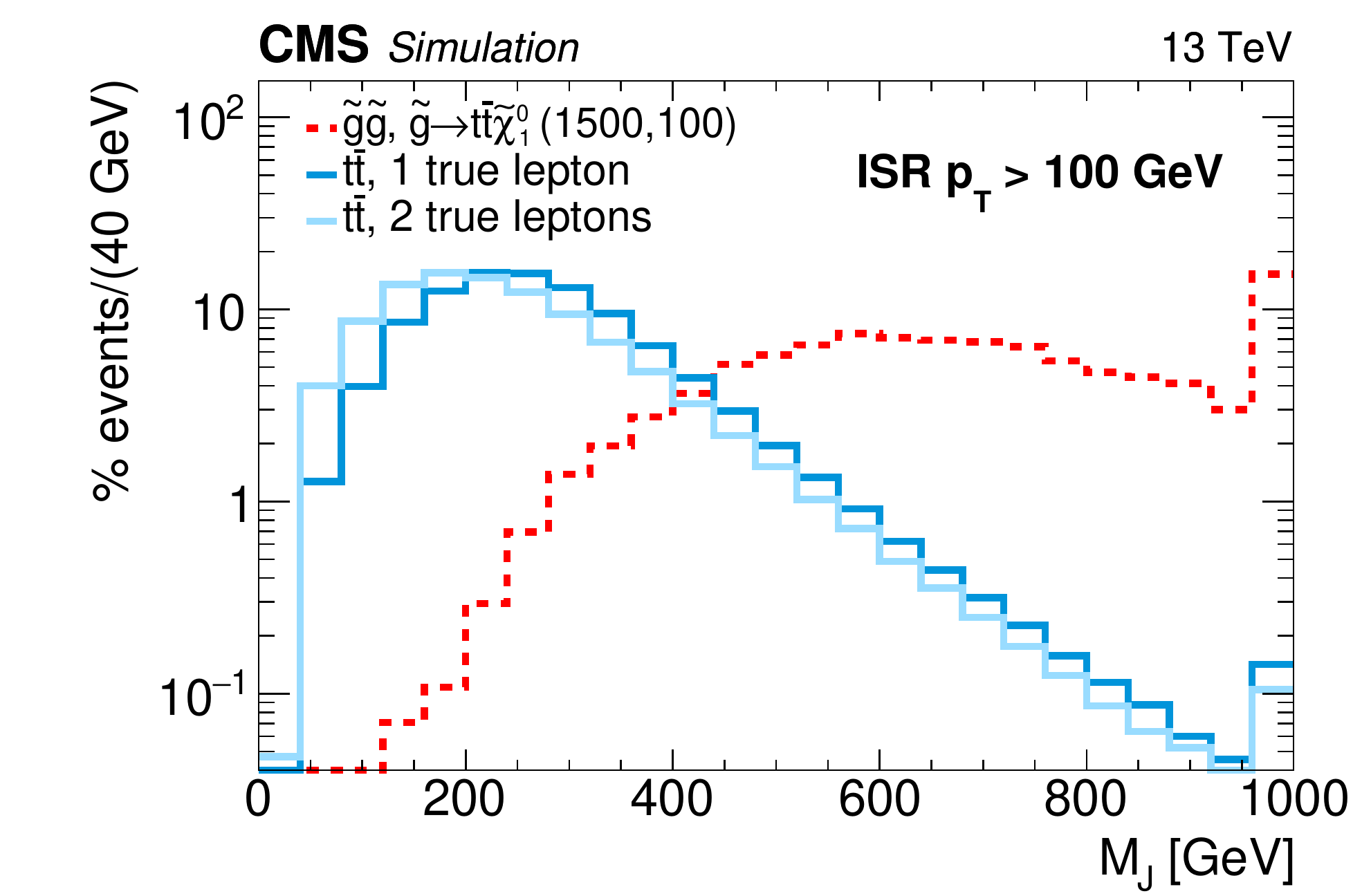}
  \caption{Distributions of \MJ, normalized to the same area, from simulated event samples
    with a small ISR contribution (left) and a significant ISR contribution (right).  These components are
    defined according to whether the $\pt$ of the $\ttbar$ system (or, in the case of signal events, that of
    the $\PSg\PSg$ system) is $<$10\GeV or $>$100\GeV, respectively. The T1tttt(NC) signal model (dashed red line),
    is described in Section~\ref{sec:samples}; the first model
    parameter in parentheses corresponds to \mGlu and the second to \mLSP, both in units of GeV.
    The events satisfy the requirements $\MET>200\GeV$ and $\HT>500\GeV$ and have at least one reconstructed lepton.
}
  \label{fig:mj_vs_isr}
\end{figure}

For \ttbar events with a small contribution from initial-state radiation (ISR), the \MJ distribution has an
approximate cutoff at twice the mass of the top quark, as shown in Fig.~\ref{fig:mj_vs_isr}
(left).  In contrast, the \MJ distribution for signal events extends to larger values.
The presence of a significant amount of ISR generates a high-$\MJ$ tail in the \ttbar
background, as shown in Fig.~\ref{fig:mj_vs_isr} (right).

The missing transverse energy, \MET, is given by the magnitude of $\ptvecmiss$,
the negative vector sum of the transverse momenta of all PF
candidates~\cite{cms-pas-pft-09-001,cms-pas-pft-10-001}.  Correspondence to the
true undetectable energy in the event is improved by replacing the contribution
of the PF candidates associated with a jet by the calibrated four-momentum of
that jet. To separate backgrounds characterized by the presence of a single
$\PW$ boson decaying leptonically but without any other source of missing energy,
the lepton and the \MET are
combined to obtain the transverse mass, \mt, defined as:
\begin{equation}
  \mt = \sqrt{2\pt^{\ell}\MET[1-\cos(\Delta\phi_{\ell, \ptvecmiss } )]},
  \label{eq:MT}
\end{equation}
where $\Delta\phi_{\ell,\ptvecmiss}$ is the difference between the
azimuthal angles of the lepton momentum vector and the missing momentum vector,
$\ptvecmiss$. Finally, we define the quantity \HT as the scalar sum of the
transverse momenta of all the \smalr jets passing the selection.

\section{Trigger and event selection}
\label{sec:eventSelection}

The data sample used in this analysis was obtained with triggers that require $\HT>350\GeV$ and at least one
electron or muon with $\pt>15\GeV$, where these variables are computed with online (trigger-level)
quantities and typically have somewhat poorer resolution than the corresponding offline variables.
To ensure high trigger efficiency with respect to the offline definition of
lepton isolation described in the previous section (mini-isolation), we designed these triggers
with very loose lepton isolation requirements and fixed the isolation cone size to $R=0.2$.
For events passing the offline selection, the total trigger efficiencies, measured in data control samples that
are independently triggered, are found to be $(95.1\pm 1.1)\%$ for the muon channel and $(94.1\pm 1.2)\%$ for
the electron channel and are independent of the analysis variables within the uncertainties.
These efficiencies are applied to the simulation as a correction.

The offline event selection is summarized in Table~\ref{tab:selection:cutflow}, which lists the event yields
expected from simulation for both SM background processes and for the two benchmark T1tttt signal models.
We select events with exactly one isolated charged lepton (an electron or a muon), $\HT>500\GeV$,
$\MET>200\GeV$, and at least six jets, at least one of which is \cPqb-tagged.
After this set of requirements, referred in the following as the {\it baseline selection}, more
than 80\% of the remaining SM background arises from \ttbar production. The contributions from events with a single
top quark or a \PW~boson in association with jets are each about 6--7\%.  The background from QCD multijet events
after the baseline selection is negligible due to the combination of leptonic, \MET, and \njets requirements.

\begin{table}[btp!]\centering
\topcaption{Event yields obtained from simulated event samples, as the event selection criteria are applied.
The category \textit{Other} includes Drell--Yan, $\ttbar\PH(\to \bbbar)$, $\ttbar\ttbar$,
$\PW\Z$, and $\PW\PW$. The yields for \ttbar events in fully hadronic final states
are included in the QCD multijet category. The category $\ttbar\mathrm{V}$ includes $\ttbar\PW$, $\ttbar\Z$, and $\ttbar\gamma$.
The benchmark signal models, T1tttt(NC) and T1tttt(C), are described in Section~\ref{sec:samples}.
The event selection requirements listed above the
horizontal line in the middle of the table are defined as the \textit{baseline selection}.
The background estimates before the $\HT$ requirement are not specified because some of the
simulated event samples do not extend to the low $\HT$ region.
Given the size of the MC samples described in Section~\ref{sec:samples}, rows with zero yield have statistical uncertainties
of at most 0.16 events, and below 0.05 events in most cases.}
\label{tab:selection:cutflow}
 \renewcommand{\arraystretch}{1.1}
\resizebox{\textwidth}{!}{
\begin{tabular}[tbp!]{ l | rrrrrrr | r | r | r }\hline
 \multicolumn{1}{c|}{${\cal L} = 2.3$\fbinv}  & Other & QCD & $\ttbar\mathrm{V}$ & Single $\cPqt$ & \wjets
                            & $\ttbar$ (1$\ell$) & $\ttbar$ ($2\ell$) & SM bkg.  & T1tttt(NC) & T1tttt(C)\\ \hline
No selection          & \multicolumn{1}{c}{---}    & \multicolumn{1}{c}{---}     & \multicolumn{1}{c}{---}   & \multicolumn{1}{c}{---}    & \multicolumn{1}{c}{---}     & \multicolumn{1}{c}{---}     & \multicolumn{1}{c|}{---}    & \multicolumn{1}{c|}{---}      & 31.3 & 190.0 \\
$1\ell$, $\pt>20\GeV$ & \multicolumn{1}{c}{---}    & \multicolumn{1}{c}{---}     & \multicolumn{1}{c}{---}   & \multicolumn{1}{c}{---}    & \multicolumn{1}{c}{---}     & \multicolumn{1}{c}{---}     & \multicolumn{1}{c|}{---}    & \multicolumn{1}{c|}{---}      & 11.9 & 68.7  \\
$\HT>500\GeV$         & 4131.9 & 31831.5 & 721.9 & 2926.6 & 31885.1 & 27628.7 & 3357.8 & 102483.4 & 11.9 & 44.9  \\
$\MET>200\GeV$        & 310.6  & 154.7   & 89.1  & 457.2  & 4343.1  & 2183.6  & 584.0  & 8122.3   & 10.5 & 21.5  \\
$\njets\geq6$, $\pt>30\GeV$         & 27.3   & 8.0     & 36.8  & 82.8   & 278.7   & 792.3   & 171.4  & 1397.4   & 9.6  & 20.4  \\
$\nb\geq1$            & 9.4    & 2.7     & 29.6  & 63.9   & 66.3    & 632.2   & 137.4  & 941.4    & 9.1  & 19.1  \\  \hline
$\MJ>250\GeV$         & 6.7    & 2.6     & 22.6  & 43.8   & 46.1    & 455.2   & 87.2   & 664.2    & 9.0  & 16.5  \\
$\mt>140\GeV$         & 0.7    & 1.4     & 3.0   & 3.5    & 1.2     & 5.5     & 32.5   & 47.9     & 7.0  & 9.2   \\
$\MJ>400\GeV$         & 0.4    & 0.8     & 1.1   & 1.4    & 0.6     & 2.8     & 9.7    & 16.7     & 6.4  & 4.5   \\
$\nb\geq2$            & 0.16   & 0.04    & 0.55  & 0.68   & 0.00    & 1.29    & 4.52   & 7.24     & 4.87 & 3.47  \\
$\MET>400\GeV$        & 0.02   & 0.00    & 0.12  & 0.31   & 0.00    & 0.07    & 0.72   & 1.24     & 3.60 & 1.48  \\
$\njets\geq9$, $\pt>30\GeV$ & 0.01   & 0.00    & 0.03  & 0.00   & 0.00    & 0.01    & 0.11   & 0.16     & 1.64 & 1.00  \\
 \hline
\end{tabular}
}
\end{table}

After the baseline selection requirements are applied, events are binned in several other kinematic variables,
both to increase the signal sensitivity and to define control regions,
as described in Section~\ref{ssec:bkgest:overview}. To illustrate the effect of additional
requirements, Table~\ref{tab:selection:cutflow} lists the expected yields for examples of event selection
requirements on \MJ, \mt, \njets, and \nb. The events satisfying the baseline selection are divided in the
$\MJ$-$\mt$ plane into a signal region, defined by the additional
requirements $\MJ>400\GeV$ and $\mt>140\GeV$, and three control samples, bounded
by $\MJ > 250\GeV$, that are used in the background estimation.
Approximately 37\% of signal T1tttt events are selected with the
single-lepton requirement only. In non-compressed
spectrum models, for which $\mGlu$ is significantly larger than $\mLSP$, more than half of the events
passing the lepton requirement lie in the signal region. For compressed spectrum
models, where $\mLSP\approx\mGlu-2\mTop$, the \MJ, \HT, and \MET spectra become much softer
and, as a result, only 5--10\% of the single-lepton signal events are selected.

\begin{figure}[tbp!]
\centering
\includegraphics[width=0.55\textwidth]{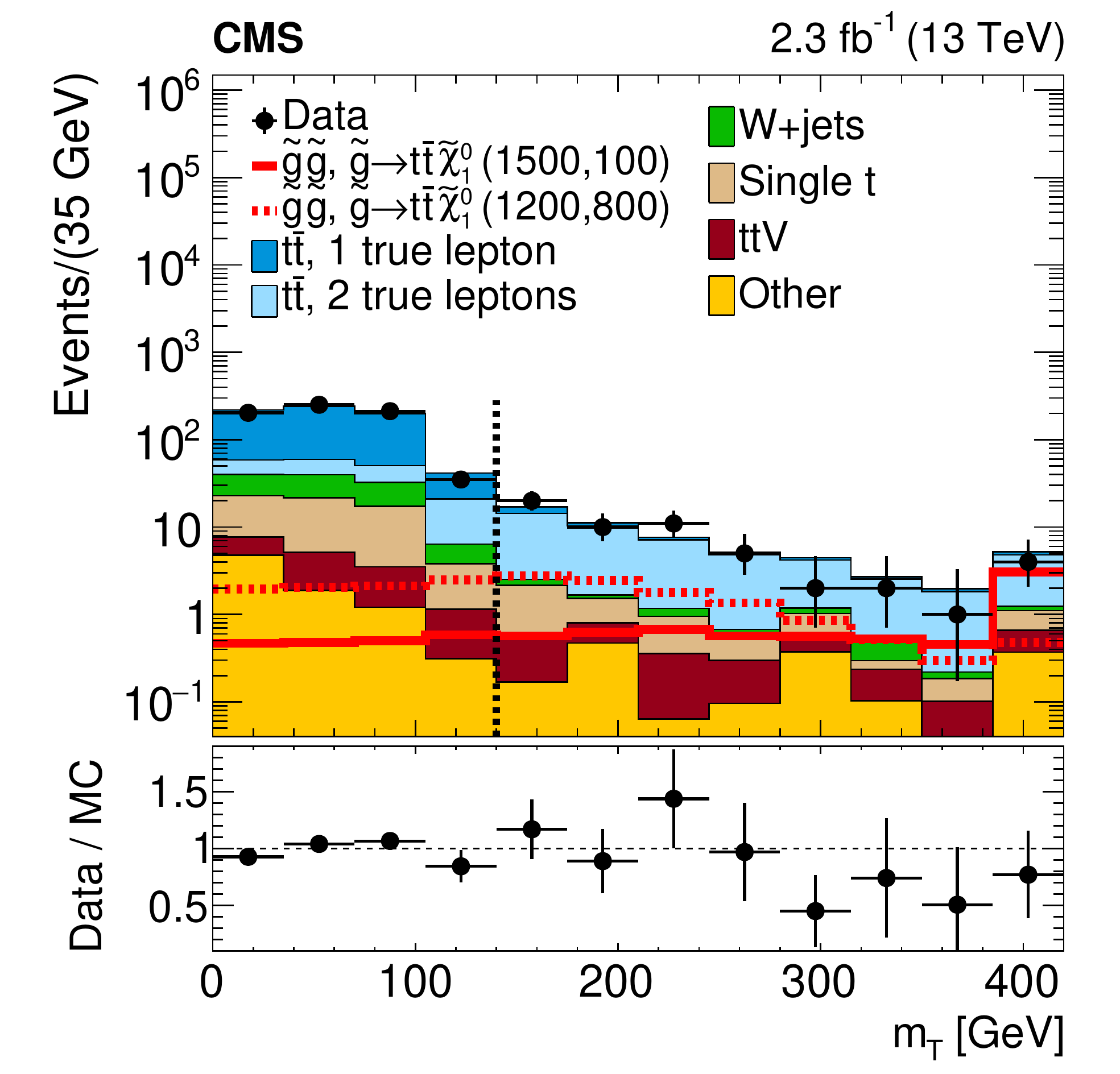}
\caption{Distribution of \mt in data and simulated event samples after the baseline selection is applied.
The background contributions shown here are from simulation, and their total yield is normalized to the number
of events observed in data.
The signal distributions are normalized to the expected cross sections.  The dashed vertical line indicates
the $\mt>140\GeV$ threshold that separates the signal regions from the control samples.}
\label{fig:selection:mt}
\end{figure}

As shown in Fig.~\ref{fig:selection:mt},
backgrounds with a single \PW~boson decaying leptonically are strongly suppressed after the $\mt>140\GeV$
requirement, so the total SM background in the signal region is dominated by dilepton \ttbar events.
This dilepton background falls into two categories, which make roughly equal contributions. The first involves an
identified electron or muon and a hadronically decaying $\tau$ from $\PW$ decay. The second source
involves two leptons, each of which is an electron or a muon. One of the leptons fails to satisfy the
lepton selection criteria, which include the $\pt$ and isolation requirements. This missed lepton can be produced
either directly or indirectly in $\PW$ decay, where in the indirect case the lepton is the daughter of a $\tau$.

\section{Background estimation}
\label{sec:backgroundEstimation}

\subsection{Method}
\label{ssec:bkgest:overview}

\begin{figure}[tbp!]
\centering
\includegraphics[width=0.6\textwidth]{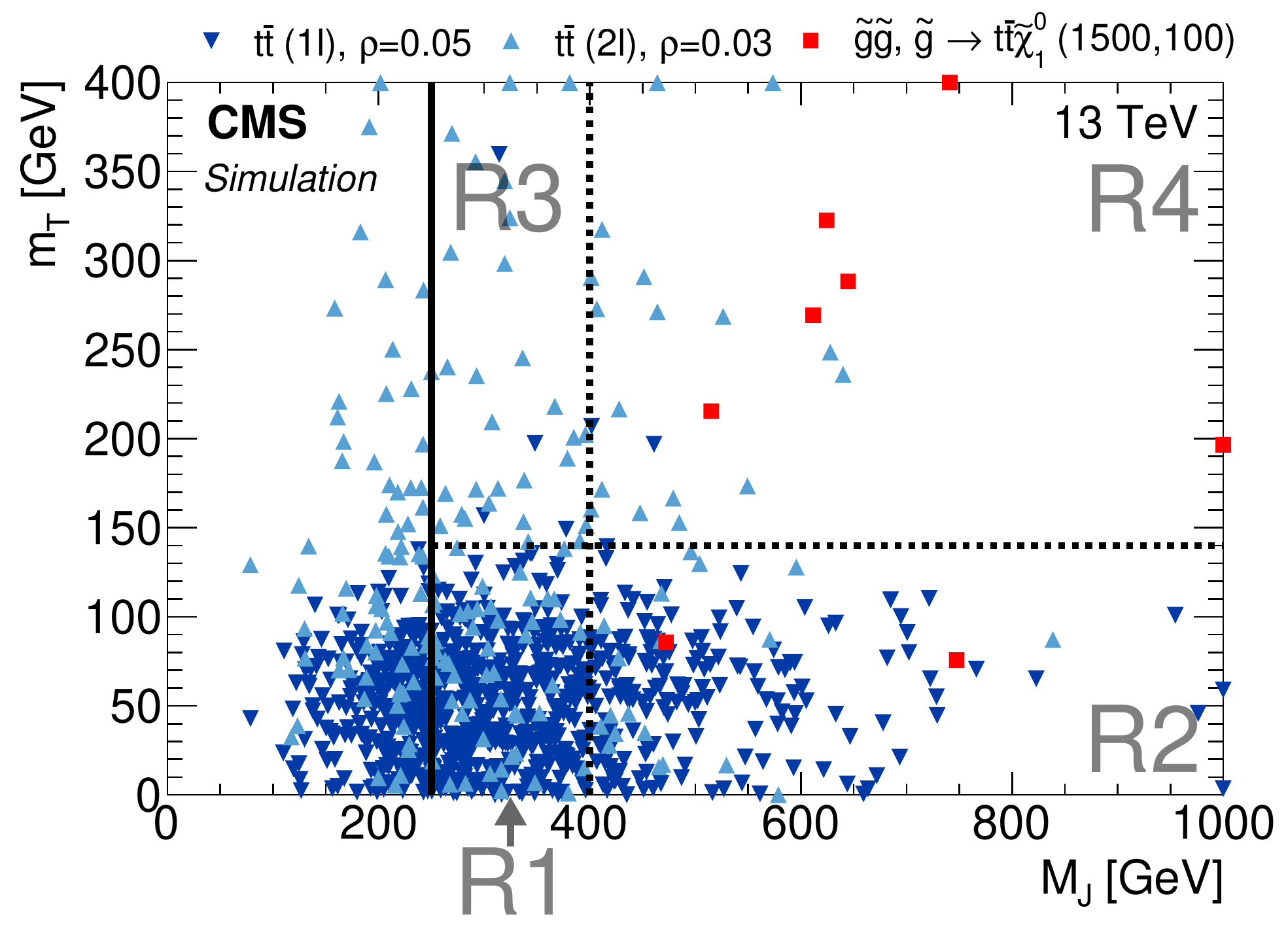}
\caption{Distribution of simulated single-lepton \ttbar events (dark-blue triangles), dilepton \ttbar events
(light-blue inverted triangles), and T1tttt(1500,100) events (red squares) in the \MJ-\mt
plane after the baseline selection. Each marker represents one expected event at 2.3\fbinv. Overflow
events are placed on the edge of the plot. The values of the
correlation coefficients $\rho$ for each background process are given in the legend. Region R4 is the nominal signal
region, while R1, R2, and R3 serve as control regions. The small signal contributions in the control regions are taken into
account in one of the global fits, as discussed in the text.}
\label{fig:selection:scat_baseline}
\end{figure}

The prediction of the background yields in each of the signal bins takes advantage of the fact that the \MJ
and \mt distributions of events with a significant amount of ISR are largely uncorrelated.
The correlation coefficients for the single-lepton and dilepton \ttbar events in the
$\MJ$-$\mt$ plane after the baseline selection (as shown in Fig.~\ref{fig:selection:scat_baseline})
are small, in the range 0.03 to 0.05. The absence of a substantial correlation allows us to measure the \MJ
distribution of the background at low \mt with good statistical precision, and extrapolate it to high \mt.
The underlying explanation for this behavior is not immediately obvious, given that low-\mt events originate mainly from
\ttbar events where only one of the top quarks decays leptonically ($1\ell$ \ttbar), while the high-\mt
regions are dominated by dilepton \ttbar events ($2\ell$ \ttbar). In particular, as shown in
Fig.~\ref{fig:mj_vs_isr} (left), in the absence of significant ISR, the dileptonic $\ttbar$ events have a softer \MJ
spectrum than single-lepton \ttbar events, simply because the reconstructed mass of a
leptonically decaying top quark does not include the undetected neutrino.

\begin{figure}
  \centering
  \includegraphics[width=0.49\textwidth]{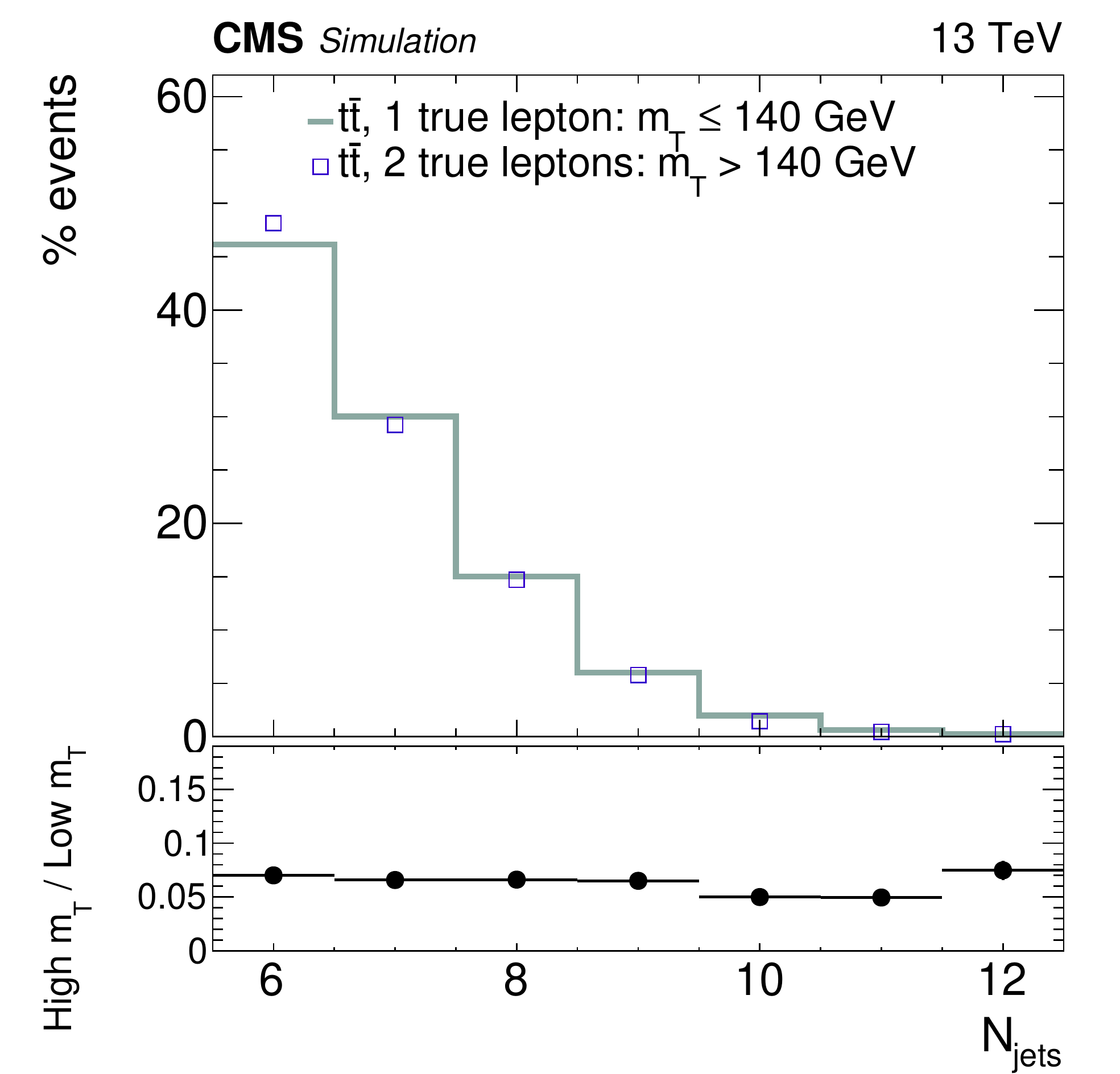}
  \includegraphics[width=0.49\textwidth]{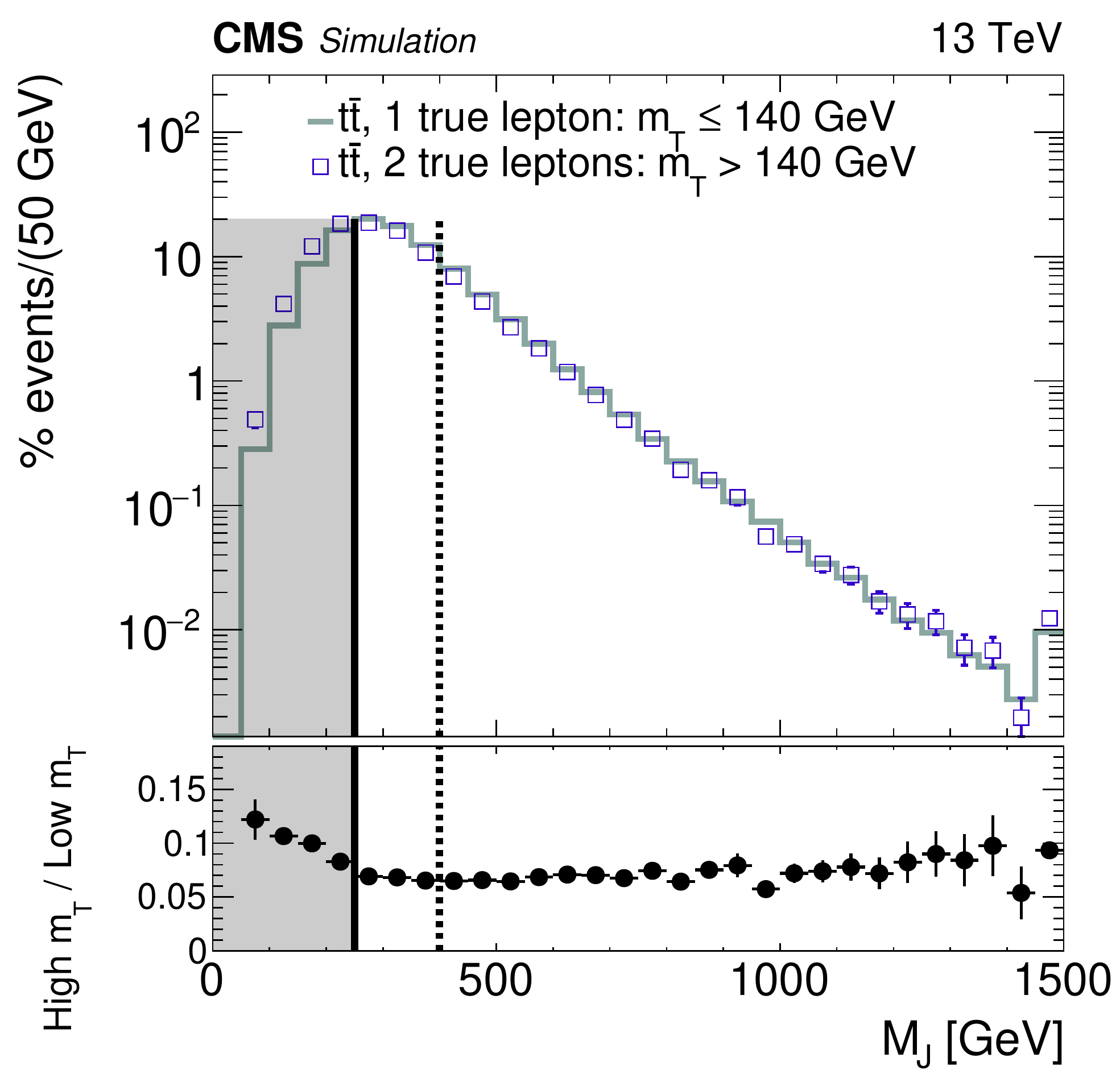}
  \caption{Comparison of \njets and \MJ distributions, normalized to the same area, in simulated \ttbar events with two true leptons at high \mt and one
    true lepton at low \mt, after the baseline selection is applied. The shapes of these distributions are similar.
    These two contributions are the dominant
    backgrounds in their respective \mt regions. The dashed vertical line on the right-hand plot indicates the $\MJ>400\GeV$
    threshold that separates the signal regions from the control samples. The shaded region corresponding to $\MJ<250\GeV$
    is not used in the background estimation.
    \label{fig:bkgest:1l2l_indep}}
\end{figure}

In events with substantial ISR, however, the contributions to \MJ from the accidental overlap of
jets can dominate the contributions due to the intrinsic mass of the top quarks.  This effect is illustrated in
Fig.~\ref{fig:bkgest:1l2l_indep}, which compares the \njets and \MJ distributions of single-lepton and dilepton
\ttbar events at high and low \mt after the baseline selection is applied. Since we require at least 6 jets, single-lepton
\ttbar events must have at least 2 ISR jets and dilepton \ttbar events must have at least 4. In this regime, the
probability of additional ISR jets is similar for events with a given number of partons of similar
momenta, and, as a result, the number of objects contributing to \MJ (jets plus the reconstructed lepton) is
comparable in $1\ell$ and $2\ell$ \ttbar events.  When these ISR jets overlap with the top quark decay products, the
masses of the resulting \largr jets are dominated by the accidental overlap and, thus, the shapes of the \MJ
distribution of $1\ell$ and $2\ell$ \ttbar events become more similar. This is the case for $\MJ>250$
GeV, where Fig.~~\ref{fig:bkgest:1l2l_indep} (right) shows that the distributions of the $1\ell$ and $2\ell$
\ttbar backgrounds have nearly the same shape, and the low-\mt to high-\mt extrapolation is warranted.

We thus divide the $\MJ$-$\mt$ plane into four regions, three control regions (CR) and one signal region (SR):
\begin{itemize}
\item Region R1 (CR): $\mt \le 140\GeV$,\, $250 \le \MJ\le 400\GeV$
\item Region R2 (CR): $\mt \le 140\GeV$,\, $\MJ> 400\GeV$
\item Region R3 (CR): $\mt>140\GeV$,\, $250 \le \MJ\le 400\GeV$
\item Region R4 (SR): $\mt>140\GeV$,\, $\MJ> 400\GeV$.
\end{itemize}
These regions are further subdivided into 10 bins of \MET, \njets, and \nb to increase signal sensitivity:
\begin{itemize}
\item Six bins with $200<\MET\leq400\GeV$: $(6\le\njets\le 8, \njets\geq 9) \times (\nb=1, \nb=2, \nb\geq3 )$
\item Four bins with $\MET>400\GeV$:  $(6\le\njets\le 8, \njets\geq 9) \times (\nb=1, \nb\geq2)$,
\end{itemize}
where the multiplication indicates that the binning is two dimensional in $\njets$ and $\nb$.
Given that the main background processes have two or fewer \PQb quarks, the total SM contribution to the
$\nb\geq3$ bins is very small and is driven by the \PQb-tag fake rate.
Signal events in the T1tttt and T5tttt models
are expected to populate primarily the bins with $\nb\geq2$, while bins with $\nb=1$
mainly serve to test the method in a background dominated region.

To obtain an estimate of the background rate in each of the signal bins, a modified version of an ``ABCD'' method
is used. Here, the symbols A, B, C, and D refer to four regions in a two-dimensional space in the data, where
one of the regions is dominated by signal and the other three by backgrounds.
In a standard ABCD method, the background rate in the signal region is estimated from the yields in
three control regions with the expression
\begin{equation}
\label{eq:bkgest:abcd}
\mu^\text{bkg}_\mathrm{R4} = N_\mathrm{R2}\, N_\mathrm{R3}/N_\mathrm{R1},
\end{equation}
where the labels on the regions correspond to those shown in Fig.~\ref{fig:selection:scat_baseline}.
The background prediction is unbiased in the limit that the
two variables that define the plane (in this case, \MJ and \mt) are uncorrelated.
The effect of any residual correlation is
corrected with factors $\kappa$ that can be obtained from simulated event samples:
\begin{equation} \label{eq:bkgext:kappa}
\kappa = \frac{N_\mathrm{R4}^\text{MC,bkg}/N_\mathrm{R3}^\text{MC,bkg}}{N_\mathrm{R2}^\text{MC,bkg}/N_\mathrm{R1}^\text{MC,bkg}}.
\end{equation}
When the two ABCD variables are uncorrelated or nearly so, the $\kappa$ factors are close to unity.
This procedure ignores potential signal contamination in the control regions, which is
accounted for by incorporating the constraints in Eqs.~\ref{eq:bkgest:abcd} and \ref{eq:bkgext:kappa}
into a fit that includes both signal and background components,
as described in Section~\ref{ssec:bkgdEstimationProcedure}.

In principle, the background in the 10 signal bins could be estimated by applying this procedure
in 10 independent planes.
However, this procedure would incur large statistical
uncertainties in some bins due to low numbers of events in R3. This problem is especially important in bins with a
high number of jets, where the \MJ spectrum shifts to higher values and the number of background events
expected in R4 can exceed the background in R3.

\begin{figure}[tbp!]
  \centering
  \includegraphics[width=1.0\textwidth]{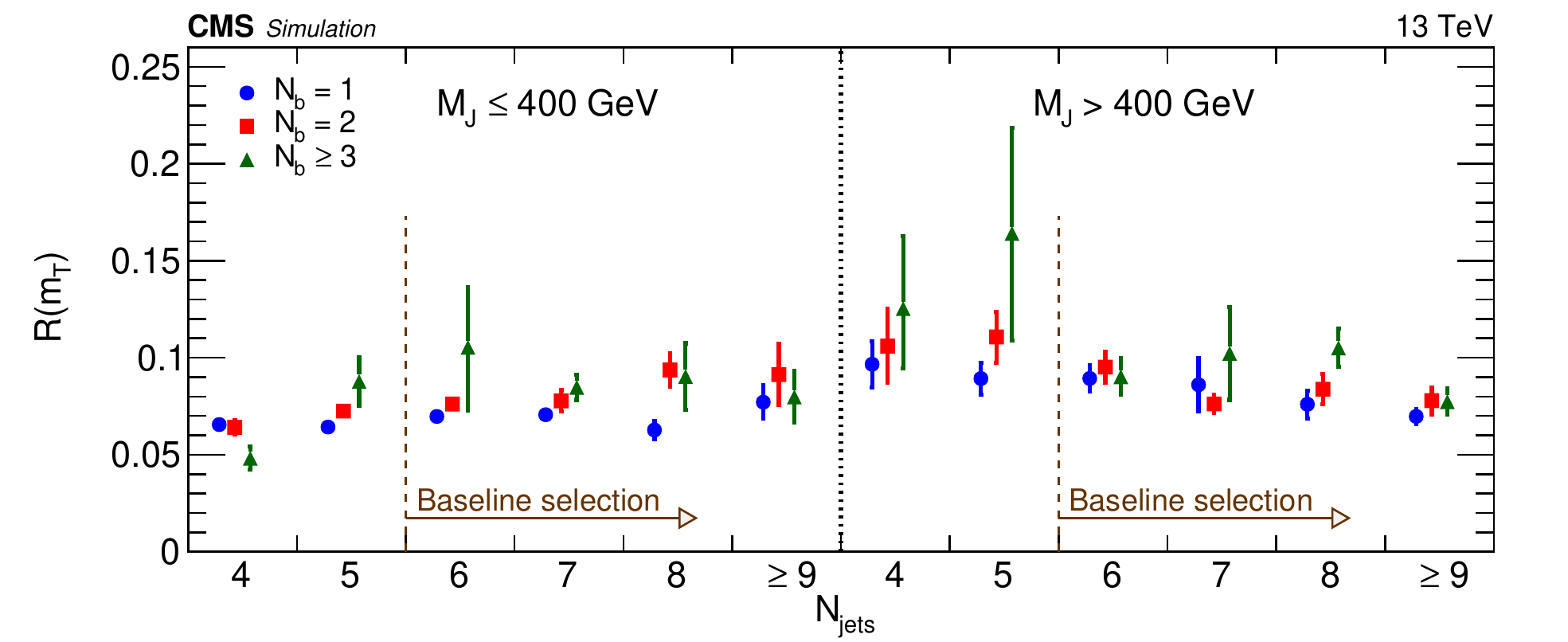}
  \caption{The ratio $R(\mt)$ of high-\mt (R3 and R4) to low-\mt (R1 and R2) event yields for the simulated SM background, as a function of
    \njets and \nb. The baseline selection requires $\njets\geq6$. The uncertainties shown are statistical only.}
  \label{fig:bkgest:rmt_allmet}
\end{figure}

\begin{figure}[tbp!]
  \includegraphics[width=0.75\textwidth]{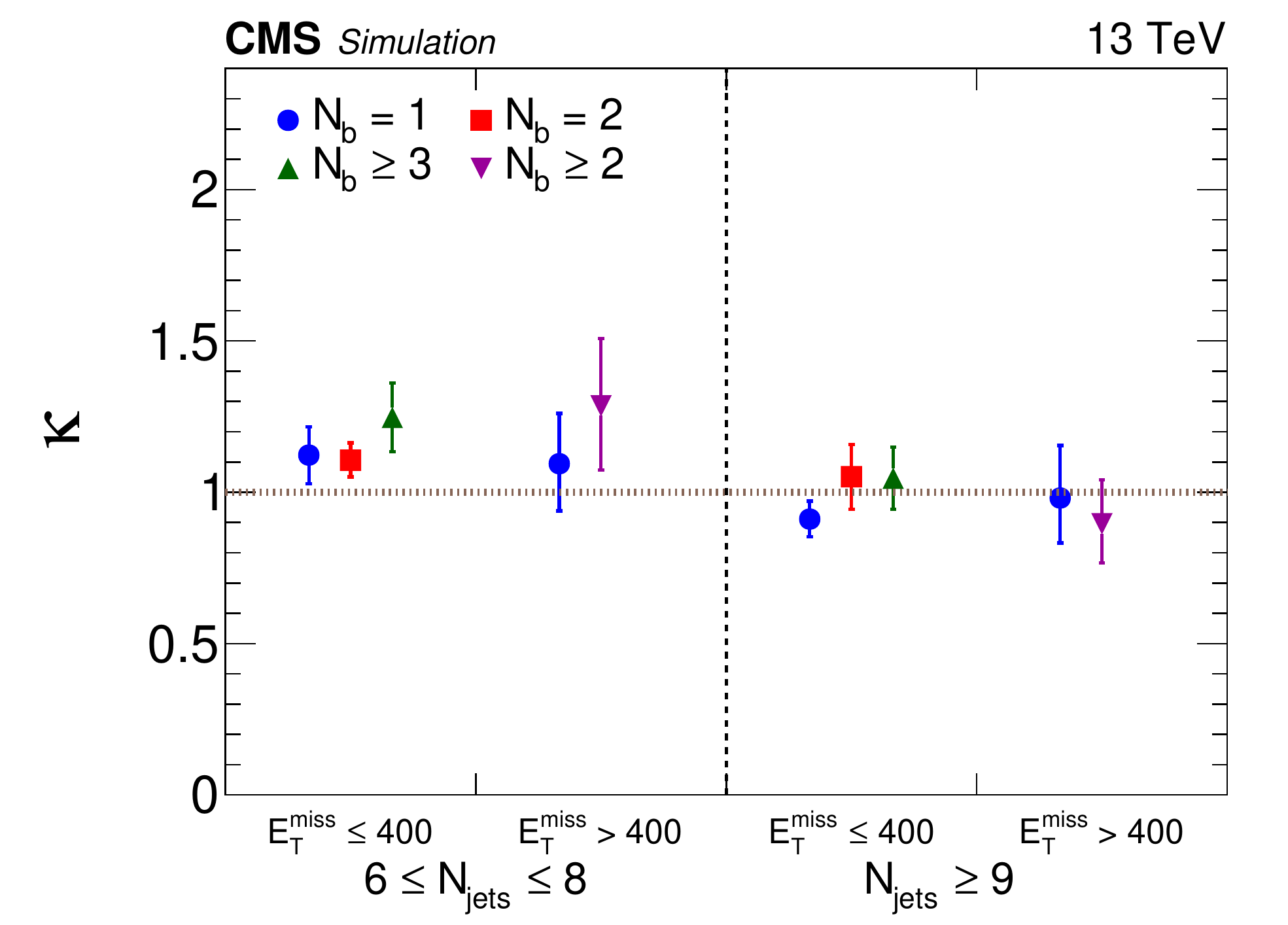}
  \centering \caption{Values of the double-ratio $\kappa$ in each of the 10 signal bins, calculated
    using the simulated SM background. The $\kappa$ factors are close to unity,
    indicating the small correlation between \MJ and \mt. The uncertainties shown are statistical only.
    \label{fig:bkgest:kappa}}
\end{figure}

To alleviate this problem, we exploit the fact that, after the baseline selection, the background is dominated
by just one source (\ttbar events), and the shapes of the \njets distributions are nearly identical
for the single-lepton and dilepton components (due to the large amounts of ISR). As a result, the \mt distribution is
approximately independent of \njets and \nb. We study this behavior with the ratio of the number of events at high to
low \mt:
\begin{equation}
  R(\mt)\equiv\frac{N(\mt>140\GeV)}{N(\mt\leq140\GeV)}.
\end{equation}
Because, as seen in Fig.~\ref{fig:bkgest:rmt_allmet}, the values of $R(\mt)$ do not vary substantially across
\njets and \nb bins, the predicted value of $R(\mt)$ is not sensitive to the modeling of the distributions of
those quantities.  We exploit this result by integrating the yields of the low-\MJ regions (R1 and R3) over
the \njets and \nb bins for each \MET bin. This procedure increases the statistical power of the ABCD method
but also introduces a correlation among the predictions (Eq.~\ref{eq:bkgest:abcd})
for the \njets and \nb bins associated with a given \MET bin.
Figure~\ref{fig:bkgest:kappa} shows the $\kappa$ factors for the 10 signal bins after summing over
\njets and \nb in R1 and R3. In all cases, their values are close to unity.

\subsection{Implementation}
\label{ssec:bkgdEstimationProcedure}

The method outlined in Section~\ref{ssec:bkgest:overview} is implemented with a likelihood function that
incorporates the statistical and systematic uncertainties in $\kappa$,
accounts for correlations arising from the common R1 and R3 yields, and corrects for
signal contamination in the control regions.

The SM background contribution for each region is described as follows.
We define $\mu_{\mathrm{R}i}^\text{bkg}$ as the estimated (Poisson) mean background in each region R$i$, with $i=1,2,3,4$.
Then, in an ABCD background calculation, these four rates can be expressed in terms
of three floating fit parameters $\mu$, $R(\mt)$, and $R(\MJ)$, and the correlation correction factor $\kappa$, as
\begin{equation}\begin{aligned}
  \mu_{\text{R1}}^{\text{bkg}}&=\mu, &
  \mu_{\text{R2}}^{\text{bkg}}&=\mu\ghmdot R(\MJ), \\
  \mu_{\text{R3}}^{\text{bkg}}&=\mu\ghmdot R(\mt),
  & \mu_{\text{R4}}^{\text{bkg}}&=\kappa\ghmdot \mu\ghmdot R(\MJ)\ghmdot R(\mt).
\end{aligned} \label{eq:rateParams}\end{equation}
Here, $\mu$ is the background rate fit parameter for R1, $R(\MJ)$ is the ratio
of the R2 to R1 rates, and $R(\mt)$ is the ratio of the R3 to R1 rates. The quantity
$\kappa$ is given by Eq.~\ref{eq:bkgext:kappa} after replacing the yields
$N_{\mathrm{R}i}^\text{MC,bkg}$ by the background rate fit parameters $\mu_{\mathrm{R}i}^\text{MC,bkg}$.

Similarly, we define $N^\text{data}_{\mathrm{R}i}$ as
the observed data yield in each region, $\mu_{\mathrm{R}i}^{\text{MC,sig}}$ as the expected signal rate in each region, and $r$ as the
parameter quantifying the signal strength relative to the expected yield across all analysis regions. We can then write the likelihood
function as
\begin{align}
  \mathcal{L}&=\mathcal{L}_\mathrm{ABCD}^\text{data}\ghmdot\mathcal{L}^\mathrm{MC}_{\kappa}\ghmdot\mathcal{L}^\mathrm{MC}_\text{sig},
  \label{eq:bkgest:likelihood}\\
  \mathcal{L}_\mathrm{ABCD}^\text{data} &= \prod_{i=1}^{4}\prod_{k=1}^{N_\text{bins}(\mathrm{R}i)}\text{Poisson}(N^\text{data}_{\mathrm{R}i,k}|\mu_{\mathrm{R}i,k}^{\text{bkg}}
  +r\ghmdot\mu_{\mathrm{R}i,k}^{\text{MC,sig}}), \label{eq:bkgest:Labcd}\\
  \mathcal{L}^\mathrm{MC}_{\kappa} &= \prod_{i=1}^{4}\prod_{k=1}^{N_\text{bins}(\mathrm{R}i)}\text{Poisson}(N_{\mathrm{R}i,k}^{\text{MC,bkg}}|\mu_{\mathrm{R}i,k}^{\text{MC,bkg}}),\\
  \mathcal{L}^\mathrm{MC}_\text{sig} &= \prod_{i=1}^{4}\prod_{k=1}^{N_\text{bins}(\mathrm{R}i)}\text{Poisson}(N_{\mathrm{R}i,k}^{\text{MC,sig}}|\mu_{\mathrm{R}i,k}^{\text{MC,sig}}). \label{eq:bkgest:Lsig}
\end{align}
The indices $k$ run over each of the \MET, \njets, and \nb bins defined in the previous section; these indices were
suppressed in Eq.~\ref{eq:rateParams} for simplicity.
Given the integration over \njets and \nb at low \MJ, $N_\text{bins}(\mathrm{R}1)=N_\text{bins}(\mathrm{R}3)=2$, while
$N_\text{bins}(\mathrm{R}2)=N_\text{bins}(\mathrm{R}4)=10$.

In Eq.~\ref{eq:bkgest:likelihood}, $\mathcal{L}_\mathrm{ABCD}^\text{data}$ accounts for the statistical
uncertainty in the observed data yield in the four ABCD regions, and $\mathcal{L}^\mathrm{MC}_{\kappa}$ and
$\mathcal{L}^\mathrm{MC}_\text{sig}$ account for the uncertainty in the computation of the $\kappa$ correction
factor and signal shape, respectively, due to the finite size of the MC samples.

The systematic uncertainties in $\kappa$ and the signal efficiency are described in the following
sections. These effects are incorporated in the likelihood function as log-normal constraints with a nuisance
parameter for each uncorrelated source of uncertainty.  These terms are not explicitly shown in the likelihood
function above for simplicity.

The likelihood function defined in Eqs.~\ref{eq:bkgest:likelihood}--\ref{eq:bkgest:Lsig} is employed in two
separate types of fits that provide complementary but compatible background estimates based on an ABCD model.
The first type of fit, which we call the \emph{predictive fit},
allows us to more easily establish the agreement of the background predictions and the observations in the
null (i.e., the background-only) hypothesis.  We do this by excluding the observations in the signal
regions in the likelihood (that is, by truncating the first product in Eq.~\ref{eq:bkgest:Labcd} at $i=3$) and
fixing the signal strength $r$ to 0. This procedure leaves as many unknowns as constraints: three \emph{data} floating
parameters ($\mu$, $R(\MJ)$, and $R(\mt)$)
and three observations ($N^\text{data}_{\mathrm{R}i,k}$ with $i=1,2,3$) for each ABCD plane.
In the likelihood function there are additional floating parameters
associated with MC quantities, which have small uncertainties.
As a result, the
estimated background rates in regions R1, R2, and R3 converge to the observed values in those bins, and we
obtain predictions for the signal regions that do not depend on the observed $N^\text{data}_{\mathrm{R}4}$.  The
predictive fit thus converges to the standard ABCD method, and the likelihood machinery becomes just a convenient
way to solve the system of equations and propagate the various uncertainties.

Additionally, we implement a \textit{global fit} which, by making use of the observations in the signal regions,
can provide an estimate of the signal strength $r,$ while allowing for signal events to populate the control
regions.  This is achieved by including all four observations, $N^\text{data}_{\mathrm{R}i,k}$ with $i=1,2,3,4$,
in the likelihood function. Since there are four observations and three floating background parameters in each ABCD
plane, there are enough constraints for the signal strength also to be determined in the fit.

\label{sec:systematicUncertainties}

\subsection{Systematic uncertainties}

This section describes the systematic uncertainties in the background prediction, which are incorporated into
the analysis as an uncertainty in the $\kappa$ correction. Because the dominant background arises from 2$\ell$
$\ttbar$ events, we use a control sample with two reconstructed leptons to validate our background estimation
procedure and to quantify the associated uncertainty.  The resulting uncertainty is augmented with
simulation-based studies of effects that are not covered by this dilepton test.
Table~\ref{tab:unc:bkgd_meth2} summarizes all of the uncertainties in the background prediction.

The ability of the ABCD method to predict the $2\ell$ \ttbar background is studied using a modified ABCD
plane, in which the high-\mt regions, R3 and R4, are replaced with regions D3 and D4, which have two reconstructed leptons.
These regions have low and high $\MJ$, respectively, just as R3 and R4. The events in D3 and D4 pass
the same selection as those in R3 and R4, except for the following changes: \njets bin boundaries are lowered
by one to keep the number of \largr jet constituents the same as in the single-lepton samples; the \mt requirement is
not applied; and events with $\nb=0$ are included to increase the size of the event sample, while events with $\nb\geq3$ are
excluded to avoid signal contamination.  We perform this test only for low $\MET$ to further avoid the potentially
large signal contribution in the high-\MET region.  The low-\MJ regions (R1 and D3) are integrated over
\njets, while the high-\MJ regions (R2 and D4) are binned in low and high \njets. The predictive fit is then
used to predict the D4 event yields for both \njets bins.
We predict $11.0\pm2.3$ ($1.5\pm0.5$) events for the low (high) \njets bin, and we observe 12 (2) events.
Given the good agreement between
prediction and observation, the statistical precision of the test is taken as a systematic uncertainty in
$\kappa$. These uncertainties are 37\% and 88\% for the low- and high-\njets regions, respectively.

Since the event composition of regions D3 and D4 is not fully representative of
that in R3 and R4, we perform studies on potential additional sources of
systematic uncertainty in the simulation.  We find that the main source of
1$\ell$ \ttbar events in the high-\mt region is jet energy mismeasurement. We
study the impact of mismodeling the size of this contribution by smearing the
jet energies by an additional 50\% with respect to the jet energy resolution
measured in data~\cite{Chatrchyan:2011ds} and calculating the corresponding shift in $\kappa$.
To ensure that there are no further
significant differences between the \MJ shapes of events reconstructed with one
or two leptons, we also calculate the shift in $\kappa$ due to jet energy
corrections, potential ISR \pt and top quark \pt mismodeling, as well as the amount of
non-\ttbar background.  Even though each of these can alone have a significant
effect on the \MJ shape, the $\kappa$ factor, as a double ratio, remains largely
unaffected (Table~\ref{tab:unc:bkgd_meth2}).  Including these uncertainties
in the likelihood fit produces a negligible contribution to the total
uncertainty.

\begin{table}[tbp!]\centering
  \topcaption{Summary of uncertainties in the background predictions. All entries in the table except for
   data sample size correspond to a relative uncertainty on $\kappa$.
   The ranges indicate the spread of each uncertainty
   across the signal bins. Uncertainties from  a particular source are treated as fully correlated across bins,
   while uncertainties from different sources are treated as uncorrelated.}
  \label{tab:unc:bkgd_meth2}
  \renewcommand{\arraystretch}{1.1}
  \begin{tabular}{ l c}\hline
    Source & Fractional uncertainty [\%] \\ \hline
    Data sample size         & \x28--118             \\
    \hline
    Dilepton control sample test   & 37--88              \\
    Simulation sample size  & \x5--17               \\
    Jet energy resolution  & \x2--10               \\
    Jet energy corrections & 1--5                \\
    ISR $\pt$              & 1--5             \\
    Top $\pt$              & 1--4             \\
    Non-\ttbar background  & \x2--11               \\
    \hline
  \end{tabular}
\end{table}

\section{Results and interpretation}
\label{sec:observationsInterpretation}

\begin{figure}[tbp!]
\centering
\includegraphics[width=0.75\textwidth]{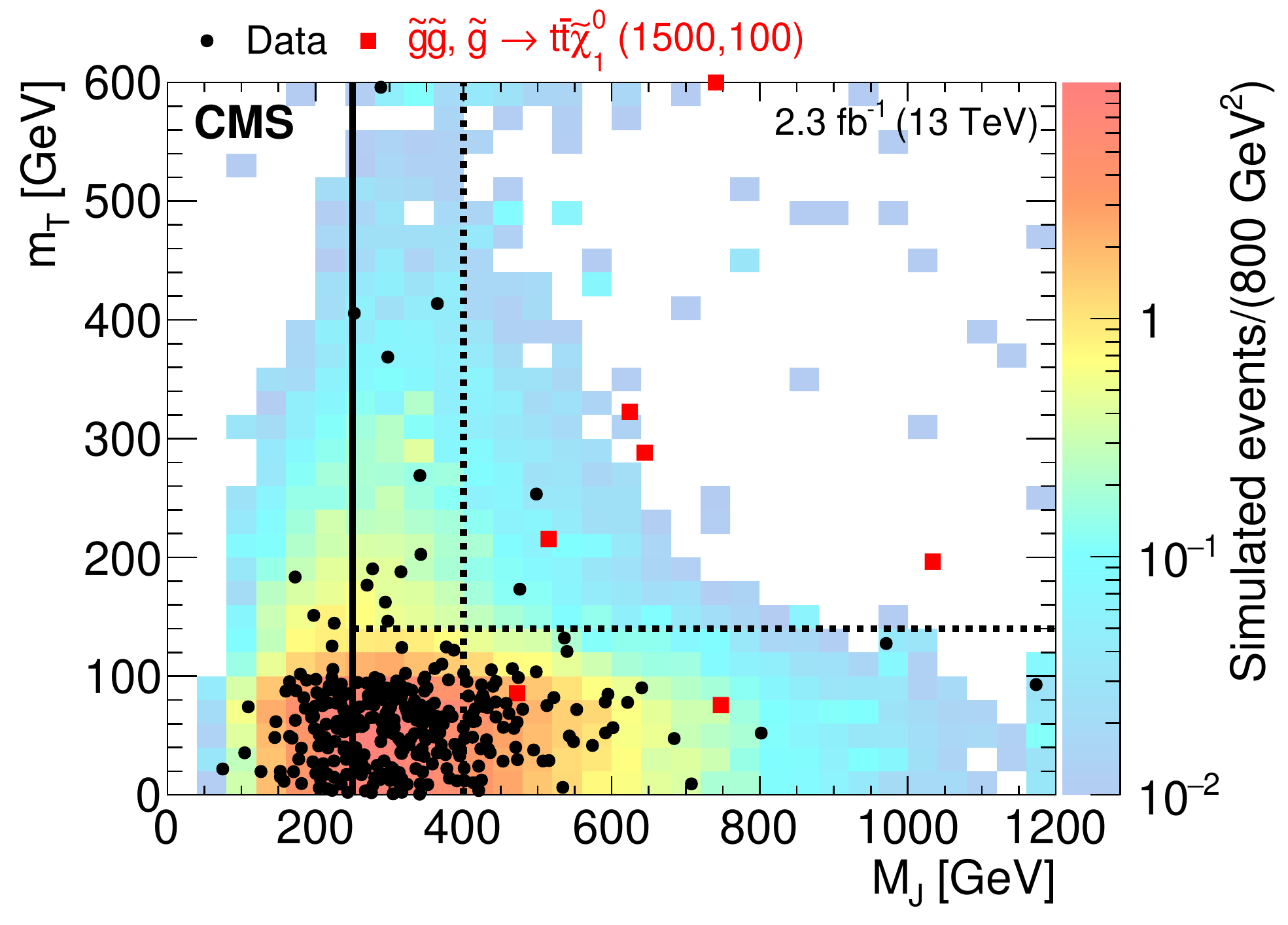}
\caption{Two-dimensional distributions for data and simulated event samples in the variables
\mt and \MJ in the $\nb\ge2$ region after the baseline selection.
The distributions integrate over the \njets and \MET bins. The black dots are the data; the colored histogram is the total
simulated background, normalized to the data; and the red dots are a particular signal sample drawn from
the expected distribution for gluino pair
production in the T1tttt model with $\mGlu=1500\GeV\text{ and }\mLSP=100\GeV$ for 2.3\fbinv.
Overflow events are shown on the edges of the plot. The definitions of the signal and control regions are the
same as those shown in Fig.~\ref{fig:selection:scat_baseline}.}
\label{fig:scatter}
\end{figure}

Figure~\ref{fig:scatter} shows the two-dimensional distribution of the data in the \mt-\MJ plane after
the baseline selection, but with the additional requirement $\nb\ge2$.  The baseline requirements include
$\MET>200\GeV$ and $\njets\ge 6$, but no further event selection is applied. For
comparison, the plot also shows the expected total SM background based on simulation,
as well as a particular sample of the expected signal distribution.
The overall distribution of events in data is consistent with the background expectation, where
the majority of events are concentrated at low \mt and \MJ. In R4, the nominal signal region, we observe only
two events in data, while, as shown in Table~\ref{tab:fit_pred_results}, the predicted SM background is about 5 events.
The T1tttt(1500,100) (NC) model would be expected to contribute 5 additional events to R4.

\begin{figure}[tbp!]
\centering
\includegraphics[width=0.49\textwidth]{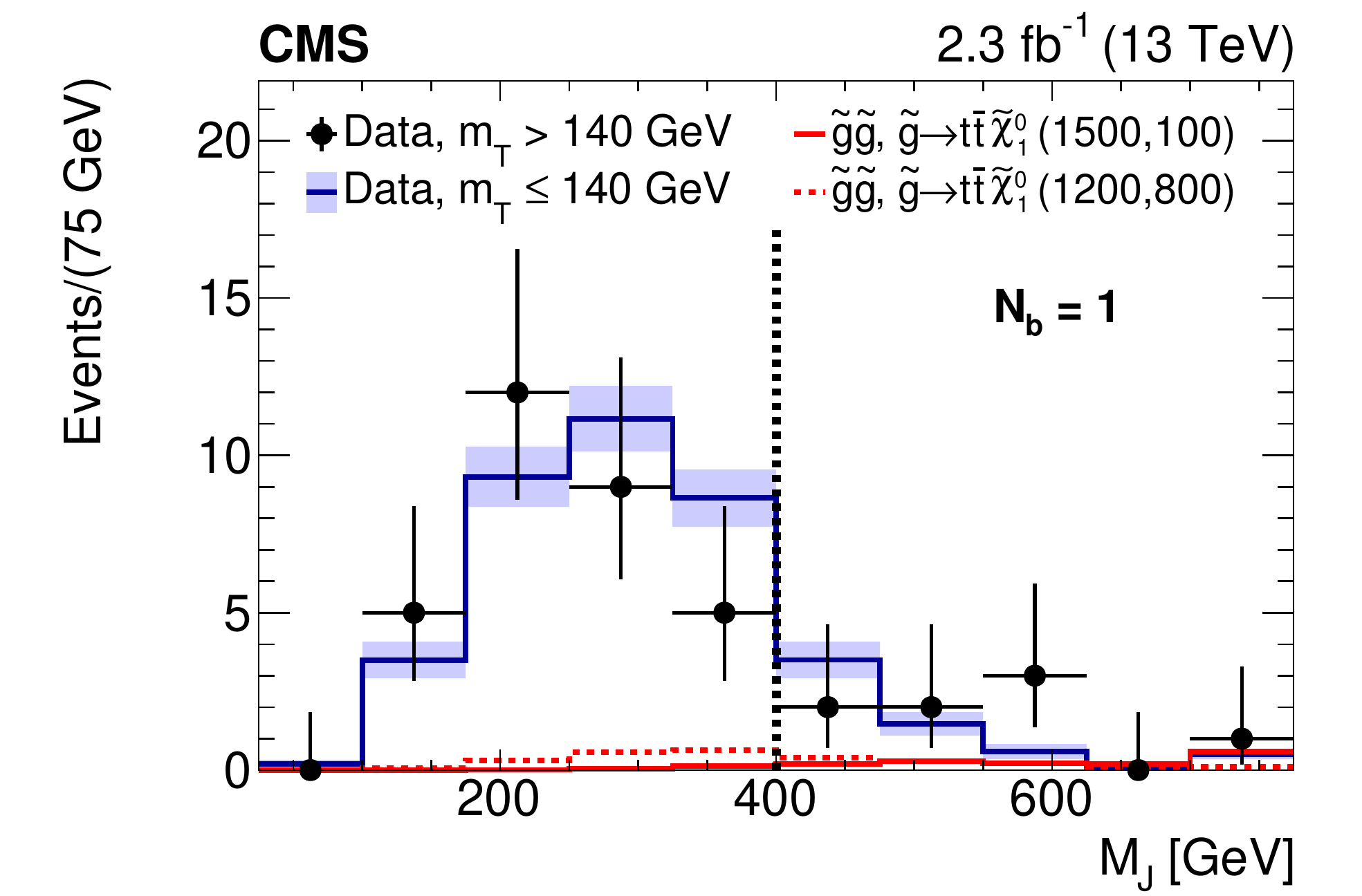}
\includegraphics[width=0.49\textwidth]{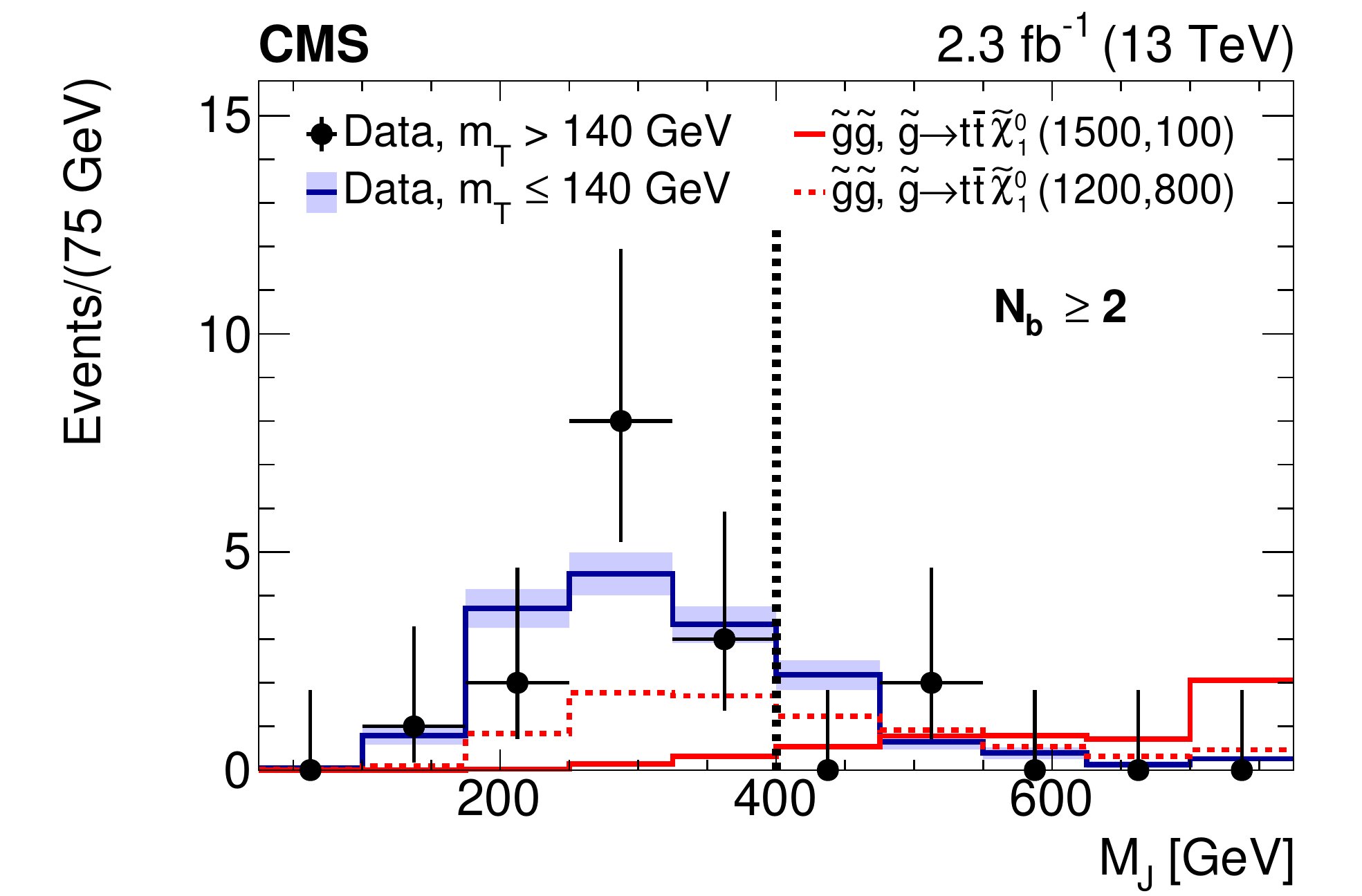}
\caption{Comparison of the \MJ distributions for low- and high-\mt in data with $\nb=1$ (left) and
$\nb\ge2$ (right) after the baseline selection.  The expected \MJ distributions of the two benchmark T1tttt scenarios for
$\mt>140~\GeV$ are overlaid.  The distributions integrate over the \njets and \MET bins. The low-\mt distribution
is normalized to the number of events in the high-\mt region.
The dashed vertical lines indicate the $\MJ>400$\GeV threshold that separates the signal regions from the control samples.}
\label{fig:data2data_mj}
\end{figure}

The validity of the central assumption of the background estimation method can be checked in the
nearly signal-free $\nb=1$ region by comparing the \MJ
shapes observed in the high- and low-\mt regions in data.
Figure~\ref{fig:data2data_mj} (left) shows the \MJ shapes in the $\nb=1$ sample,
integrating over the \njets and \MET bins. The
low \mt data have been normalized to the overall yields in the corresponding high-\mt data. The shapes of the
\MJ distributions for the high- and low-\mt regions are consistent.
Figure~\ref{fig:data2data_mj} (right) shows that the corresponding distributions in the $\nb\ge2$ sample are also
consistent, as expected in the absence of signal.

Table~\ref{tab:fit_pred_results} summarizes the observed event yields, the fitted backgrounds, and the
expected signal yields for the two T1tttt benchmark model points. Two background estimates are given:
the predictive fit (PF), which uses only the yields in regions R1, R2, and R3, and
the global fit (GF), which also incorporates region R4, as described in Section~\ref{sec:backgroundEstimation}.
In both versions of the fit, the signal strength $r$ is fixed to zero, giving results that are model
independent. (When setting limits on individual models, we allow $r$ to float, as discussed below.)
The rows labeled R4 give the results for each of the ten signal regions, as well as the corresponding
$\kappa$ factors.

In the absence of signal, the predictive fit and the version of the global fit performed under the null hypothesis, $r=0$,
should be consistent with each other.
However, because the global fit incorporates more information, specifically the yields in R4,
this fit has a smaller uncertainty. The regions with $\nb=1$ have small expected contributions from signal.
Summing over all four such signal regions (R4),
the number of estimated background events from the PF and GF
are $6.1\pm2.2$ and $5.5\pm1.3$, respectively, compared with 8 events observed in data.
The consistency between the two predictions and between the predicted and observed yields
in the R4 regions with $\nb=1$, where the signal contribution is expected to be small, serves
as a further check on the background estimation method.
Summing the yields over the six signal bins with $\nb\geq 2$, the number of estimated background
events from PF and GF is $5.6\pm 1.6$ and $4.9\pm 1.0$, respectively. In
data, we observe 2 events, lower than, but consistent with the background-only hypothesis.

\begin{table}[tbp!]
\centering
\topcaption{Observed and predicted event yields for the signal regions (R4) and background regions
(R1--R3) in data (2.3\fbinv). Expected yields for the two SUSY T1tttt benchmark scenarios are also given.
The results from two types of fits are reported: the predictive fit (PF) and the version of the global fit (GF)
performed under the assumption of the null hypothesis ($r=0$). The predictive fit uses the observed yields in regions
R1, R2, and R3 only and is effectively just a propagation of uncertainties. The global fit uses all
four regions. The values of $\kappa$ obtained from the simulation
fit are also listed. The first uncertainty in $\kappa$ is statistical,
while the second corresponds to the total systematic uncertainty.
The benchmark signal models, T1tttt(NC) and T1tttt(C), are described in Section~\ref{sec:samples}. } \label{tab:fit_pred_results}
\renewcommand{\arraystretch}{1.1}
\resizebox{\textwidth}{!}{
\begin{tabular}{lcrrccr}
\hline
Region: bin & $\kappa$  & T1tttt(NC)  & T1tttt(C) & Fitted $\mu^{\text{bkg}}$~(PF) & Fitted $\mu^{\text{bkg}}$~(GF) & Obs. \\
\hline\hline
\multicolumn{7}{c}{$200<\MET\leq 400\GeV$} \\ \hline
R1: all $\njets,\nb$                & ---                     & 0.1 & 3.2 & $336.0\pm18.3\x$ & $335.3\pm18.2\x$ & 336 \\
R2: $6\leq \njets\leq8$, $\nb=1$    & ---                     & 0.1 & 0.2 & $47.1\pm6.9\x$   & $49.5\pm6.9\x$   & 47  \\
R2: $\njets\geq9$, $\nb=1$          & ---                     & 0.1 & 0.3 & $7.0\pm2.6$    & $7.5\pm2.7$    & 7   \\
R2: $6\leq \njets\leq8$, $\nb=2$    & ---                     & 0.1 & 0.3 & $42.0\pm6.5\x$   & $41.1\pm6.2\x$   & 42  \\
R2: $\njets\geq9$, $\nb=2$          & ---                     & 0.1 & 0.5 & $7.0\pm2.6$    & $6.6\pm2.5$    & 7   \\
R2: $6\leq \njets\leq8$, $\nb\geq3$ & ---                     & 0.1 & 0.2 & $12.0\pm3.5\x$   & $11.1\pm3.2\x$   & 12  \\
R2: $\njets\geq9$, $\nb\geq3$       & ---                     & 0.2 & 0.6 & $1.0\pm1.0$    & $0.9\pm0.9$    & 1   \\
R3: all $\njets,\nb$                & ---                     & 0.2 & 3.8 & $21.0\pm4.6\x$   & $21.6\pm4.2\x$   & 21  \\
\hline
R4: $6\leq \njets\leq8$, $\nb=1$ & $1.12\pm 0.09\pm 0.43$ & 0.2 & 0.2 & $3.3\pm1.4$    & $3.6\pm1.0$    & 6   \\
R4: $\njets\geq9$, $\nb=1$          & $0.91\pm 0.06\pm 0.81$ & 0.2 & 0.4 & $0.4\pm0.3$    & $0.4\pm0.2$    & 1   \\
R4: $6\leq \njets\leq8$, $\nb=2$    & $1.11\pm 0.06\pm 0.42$ & 0.3 & 0.4 & $2.9\pm1.2$    & $2.9\pm0.8$    & 2   \\
R4: $\njets\geq9$, $\nb=2$          & $1.05\pm 0.11\pm 0.94$ & 0.3 & 0.6 & $0.5\pm0.3$    & $0.4\pm0.2$    & 0   \\
R4: $6\leq \njets\leq8$, $\nb\geq3$ & $1.25\pm 0.11\pm 0.47$ & 0.3 & 0.3 & $0.9\pm0.4$    & $0.9\pm0.3$    & 0   \\
R4: $\njets\geq9$, $\nb\geq3$       & $1.05\pm 0.10\pm 0.93$ & 0.3 & 0.7 & $0.1\pm0.1$    & $0.1\pm0.1$    & 0   \\
\hline\hline
\multicolumn{7}{c}{$\MET>400\GeV$}                                                                      \\ \hline
R1: all $\njets,\nb$                & ---                     & 0.1 & 0.5 & $16.0\pm4.0\x$   & $17.1\pm4.0\x$   & 16  \\
R2: $6\leq \njets\leq8$, $\nb=1$    & ---                     & 0.2 & 0.1 & $8.0\pm2.8$    & $6.8\pm2.5$    & 8   \\
R2: $\njets\geq9$, $\nb=1$          & ---                     & 0.1 & 0.2 & $1.0\pm1.0$    & $1.7\pm1.2$    & 1   \\
R2: $6\leq \njets\leq8$, $\nb\geq2$ & ---                     & 0.5 & 0.3 & $3.0\pm1.7$    & $2.5\pm1.4$    & 3   \\
R2: $\njets\geq9$, $\nb\geq2$       & ---                     & 0.4 & 0.6 & $1.0\pm1.0$    & $0.9\pm0.9$    & 1   \\
R3: all $\njets,\nb$                & ---                     & 0.4 & 0.9 & $4.0\pm2.0$    & $2.9\pm1.4$    & 4   \\
\hline
R4: $6\leq \njets\leq8$, $\nb=1$    & $1.09\pm 0.16\pm 0.42$ & 0.7 & 0.2 & $2.2\pm1.7$    & $1.2\pm0.7$    & 0   \\
R4: $\njets\geq9$, $\nb=1$          & $0.98\pm 0.16\pm 0.87$ & 0.4 & 0.3 & $0.2\pm0.3$    & $0.3\pm0.2$    & 1   \\
R4: $6\leq \njets\leq8$, $\nb\geq2$ & $1.29\pm 0.22\pm 0.50$ & 1.9 & 0.5 & $1.0\pm0.8$    & $0.5\pm0.4$    & 0   \\
R4: $\njets\geq9$, $\nb\geq2$       & $0.90\pm 0.14\pm 0.80$ & 1.6 & 1.0 & $0.2\pm0.3$    & $0.1\pm0.1$    & 0   \\
\hline
\end{tabular}
}
\end{table}

Given the absence of any significant excess, the results are interpreted first as exclusion limits on the
production cross section for T1tttt model points as a function of $\mGlu$ and $\mLSP$.
Table~\ref{tab:unc:sig} shows the ranges for the systematic uncertainties associated with predictions for the
expected signal yields, including those on the signal efficiency.
The largest uncertainties arise from the jet energy corrections and from the modeling
of ISR.  These uncertainties are generally in the range 10--20\% but can increase to $\sim$30\% as the mass
splitting between the gluino and LSP decreases~\cite{Khachatryan:2133129}.  The uncertainty associated with
the renormalization and factorization scales is determined by varying the scales independently up and down by
a factor of two; these are applied only as an uncertainty in the signal shape, i.e., the cross section is held
constant. The uncertainty associated with the b tagging efficiency is in the range 1--15\%. Uncertainties due
to pileup, luminosity~\cite{CMS-PAS-LUM-15-001}, lepton selection, and trigger efficiency are found to be
$\leq5$\%. Uncertainties for each particular source are treated as fully correlated across bins.

A 95\% confidence level (CL) upper limit on the production cross section is estimated using the modified
frequentist \cls method~\cite{Junk:1999kv,0954-3899-28-10-313,CMS-NOTE-2011-005}, with a one-sided profile
likelihood ratio test statistic. For this test, we perform the global fit under the background-only and
background-plus-signal ($r$ floating) hypotheses.  The statistical uncertainties from data counts in the
control regions are modeled by the Poisson terms in Eq.~\ref{eq:bkgest:Labcd}. All systematic uncertainties
are multiplicative and are treated as log-normal distributions.  Exclusion limits are also estimated
for $\pm1\sigma$ variations on the production cross section based on the NLO+NLL calculation~\cite{Borschensky:2014cia}.

\begin{figure}[tbp!]
\centering
\includegraphics[width=0.75\textwidth]{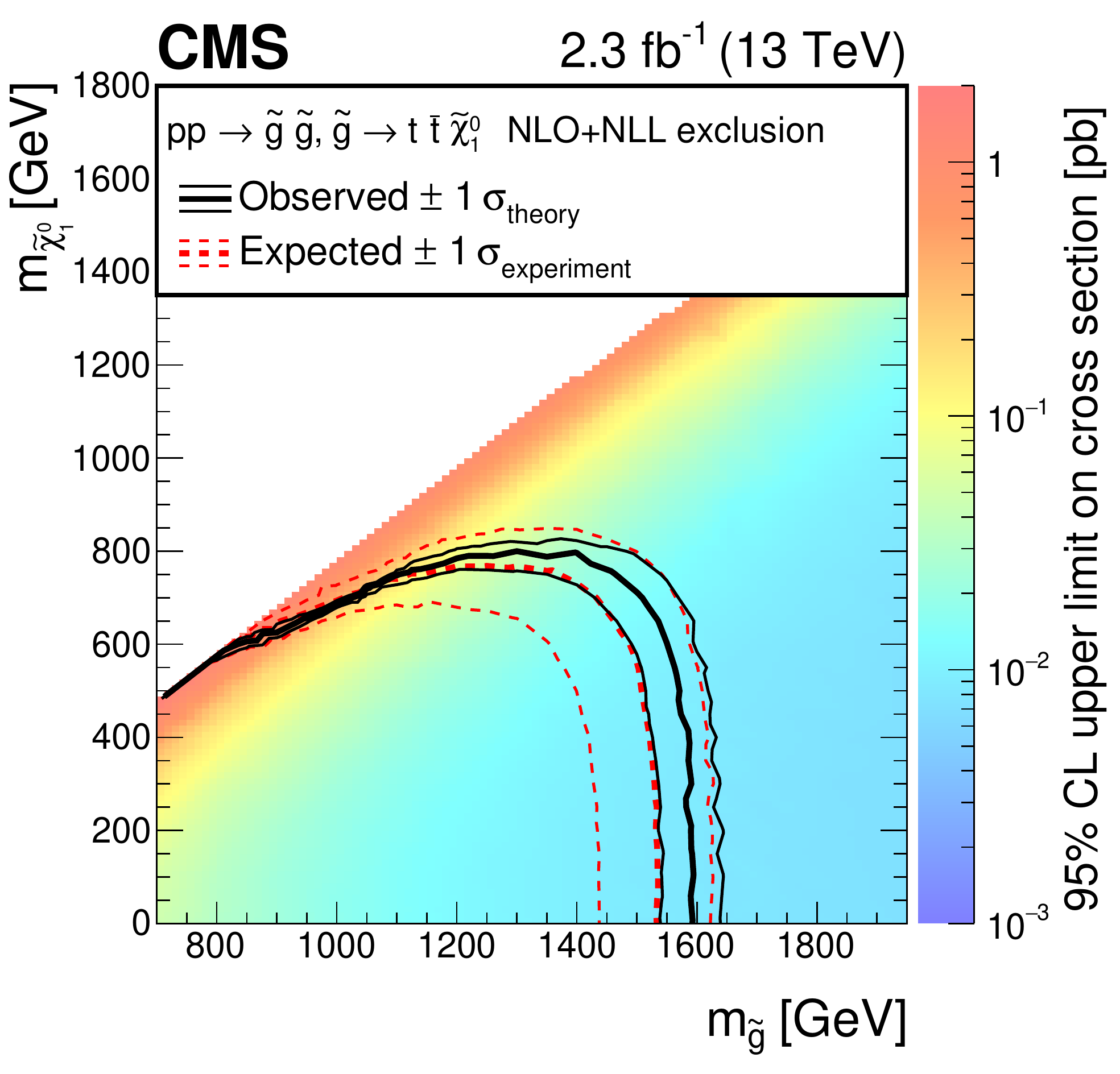} \\
\caption{Interpretation of results in the T1tttt model. The colored regions show the
upper limits (95\% CL) on the production cross section for
 $\Pp\Pp\to\PSg\PSg,\PSg\to\ttbar\PSGczDo$ in the $\mGlu$-$\mLSP$ plane. The curves
show the expected and observed limits on the corresponding SUSY particle masses obtained by
comparing the excluded cross section with theoretical cross sections.
}
\label{fig:limits_scan}
\end{figure}

Figure~\ref{fig:limits_scan} shows the corresponding excluded region at a 95\% CL for the T1tttt model in the $\mGlu
- \mLSP$ plane. At low \PSGczDo mass we exclude gluinos with masses of up to 1600\GeV. The highest limit on
the \PSGczDo mass is 800\GeV, attained for $\mGlu$ of approximately 1300~\GeV. The observed limits are within the
$1\sigma$ uncertainty in the expected limits.  The central value is slightly higher because the observed event
yield is less than the SM background prediction, as shown in Table~\ref{tab:fit_pred_results}.

In the context of natural SUSY models, it is important to extend the interpretation to scenarios in which the
top squark is lighter than the gluino. Rather than considering a large set of models with independently
varying top squark masses, we consider the extreme case in which the top squark has approximately the smallest
mass consistent with two-body decay, $\mStop \approx \mTop + \mLSP$, for a range of gluino and neutralino
masses.  The decay kinematics for such extreme, compressed mass spectrum models correspond to the lowest
signal efficiency for given values of $\mGlu$ and $\mLSP$, because the top quark and the $\PSGczDo$ are produced
at rest in the top squark frame. As a consequence, the excluded signal cross section for fixed values of
$\mGlu$ and $\mLSP$ and with $\mGlu>\mStop\ge \mTop + \mLSP$ is minimized around this extreme model point. For
physical consistency, the signal model used in this study, both in the fit procedure and in the theoretical
cross section used to obtain mass limits, includes not only gluino-pair production, but also direct
$\PSQt_1\PASQt_1$ production.  However, the effect of the direct top squark contribution on the
results is small, $\lesssim 2\%$ for $\mLSP > 400 \GeV$ and up to 20\% for low values of $\mLSP$.

\begin{figure}[tbp!]
\centering
\includegraphics[width=0.75\textwidth]{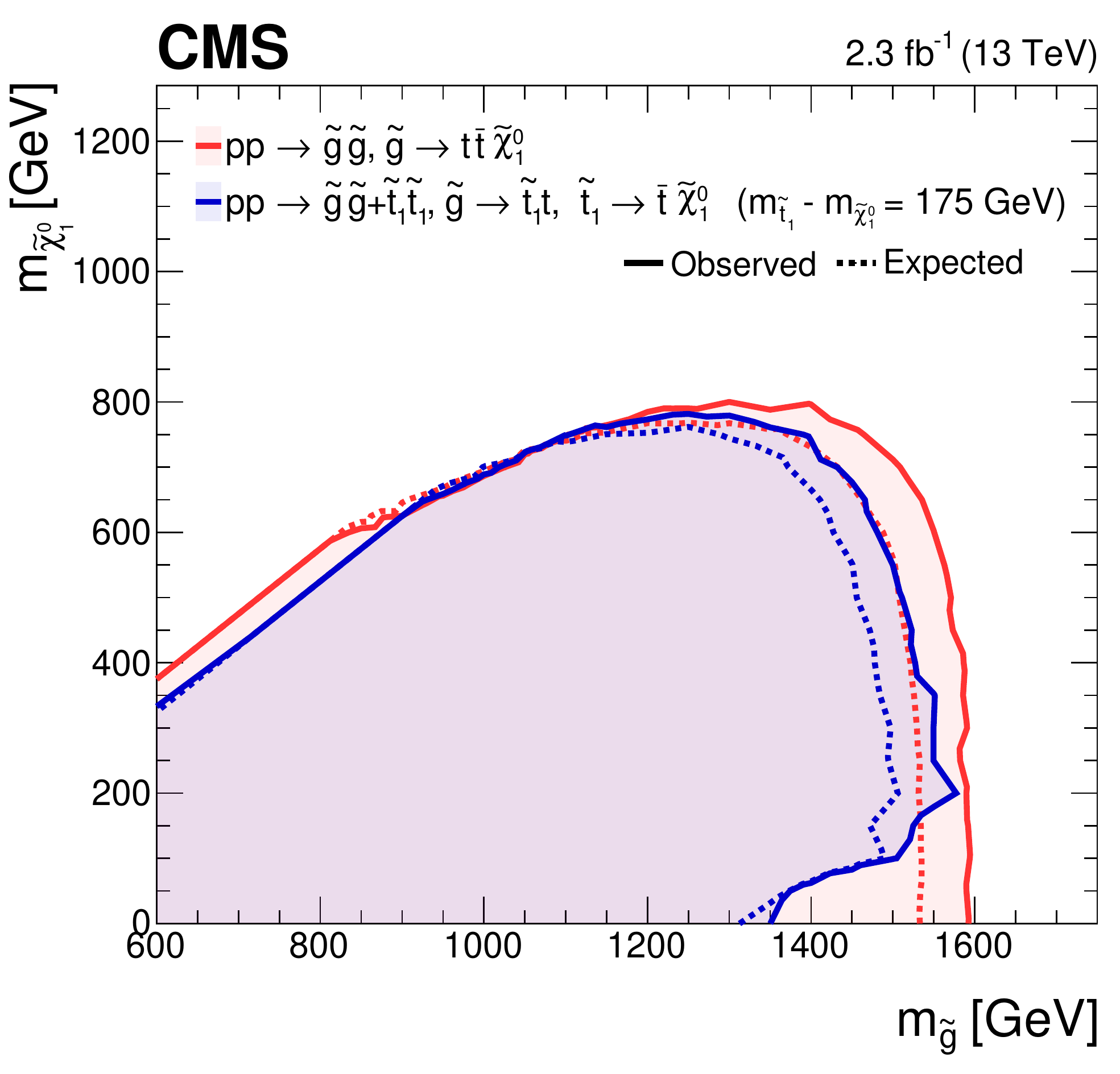} \\
\caption{Excluded region (95\% CL), shown in blue, in the $\mGlu$-$\mLSP$ plane for a model
combining T5tttt, gluino pair production,
followed by gluino decay to an on-shell top squark,
together with a model for direct top squark pair production.
The top squarks decay via the two-body process $\PSQt\to\PQt\PSGczDo$. The neutralino
and top squark masses are related by the constraint $\mStop = \mLSP + 175\GeV$. For comparison, the excluded
region (95\% CL) from Fig.~\ref{fig:limits_scan} for the T1tttt model, which has three-body gluino decay,
is shown in red. The small difference between the two boundary curves shows that the limits for the scenarios
with two-body gluino decay have only a weak dependence on the top squark mass.}
\label{fig:T5tttt_limits_scan}
\end{figure}

Figure~\ref{fig:T5tttt_limits_scan} shows the excluded region in the $\mGlu$-$\mLSP$ plane for this
combined model with both gluino-mediated top squark production and direct top squark pair production. The top
squark mass is assumed to be 175\GeV above that of the neutralino.  For most of the excluded region, the
boundary is close to that obtained for the T1tttt model, showing that there is only a weak sensitivity to the
value of the top squark mass.  The uncertainty on the boundary of the excluded region for the T5tttt model
is similar to that shown for the T1tttt model in Fig.~\ref{fig:limits_scan}.
For $\mLSP>150$ GeV, the excluded value of \mGlu is typically within 60 GeV of that excluded for T1tttt.
Models that have low values of $\mLSP$ show a reduced sensitivity because the
neutralino carries very little momentum, reducing the value of $\mt$. In this kinematic region, the
sensitivity to the signal is dominated by the events that have at least two leptonic $\PW$ boson decays, which
produce additional $\MET$, as well as a tail in the $\mt$ distribution. Although such dilepton events are
nominally excluded in the analysis, a significant number of these signal events escape the dilepton
veto. These events include both $\PW$ decays to $\tau$ leptons that decay hadronically, and $\PW$ decays to
electrons or muons that are below kinematic thresholds or are outside of the detector acceptance.

\begin{table}[tbp!]\centering
\topcaption{Typical values of the signal-related systematic uncertainties.
Uncertainties due to a particular source are treated as fully correlated between bins,
while uncertainties due to different sources are treated as uncorrelated.}
\label{tab:unc:sig}
\renewcommand{\arraystretch}{1.1}
\begin{tabular}[tbp!]{lc}
\hline
Source                     & Fractional uncertainty [\%] \\
\hline
Lepton efficiency          & 1--5    \\
Trigger efficiency         & 1     \\
b tagging efficiency           & \x1--15\\
Jet energy corrections     & \x1--30 \\
Renormalization and factorization scales      & 1--5\\
Initial state radiation    & \x1--35\\
Pileup                    & 5     \\
Integrated luminosity                 & 3     \\
\hline
\end{tabular}
\end{table}

\section{Summary}
\label{sec:summary}

Using a sample of proton-proton collisions at $\sqrt{s} = 13\TeV$ with an integrated luminosity of 2.3\fbinv,
a search for supersymmetry is performed in the
final state with a single lepton, b-tagged jets, and large missing transverse momentum.
The search focuses on final states resulting from the pair production of gluinos, which
subsequently decay via $\PSg\to\ttbar\PSGczDo$, leading to high jet multiplicities.

A key feature of the analysis is the use of the variable $\MJ$, the sum of the masses of large-$R$ jets,
which are formed by clustering anti-$\kt$ $R=0.4$ jets and leptons.
Used in conjunction with the variable $\mt$, the transverse mass
of the system consisting of the lepton and the missing transverse momentum vector,
$\MJ$ provides a powerful background estimation method that is well suited
to this high jet multiplicity search.

After the baseline selection is applied, signal (R4) and control regions (R1, R2, and R3) are defined in the $\MJ$-$\mt$
plane, which are further divided into bins of $\MET$, $\njets$, and $\nb$ to provide additional sensitivity.
In regions R3 and R4, the requirement
$\mt>140$\GeV provides strong suppression of the single-lepton $\ttbar$ background,
so that dilepton $\ttbar$ events dominate over all other background sources.
For these dilepton events to enter a signal region, however, they  must contain a substantial
amount of initial-state radiation (ISR). For this extreme range of ISR jet momentum and multiplicity,
the single-lepton and dilepton $\ttbar$ events have very similar kinematic properties. The variables
$\MJ$ and $\mt$ are nearly uncorrelated, even though different processes dominate
the low- and high-$\mt$ regions. As a consequence, the low-$\mt$
regions (R1 and R2) can be used to measure the background shape for the $\MJ$ distribution
at high $\mt$. A correction factor, near unity, is taken from simulation and is used to account for
a possible correlation between $\MJ$ and $\mt$.

The observed event yields in the signal regions are consistent with the predictions for the SM
background contributions, and exclusion limits are set on the gluino pair production
cross sections in the $\mGlu$-$\mLSP$ plane, as described by the simplified models T1tttt and T5tttt, where the latter is
augmented with a model of direct top squark pair production for consistency.
In the T1tttt model, gluinos decay via the three-body process $\PSg\to\ttbar\PSGczDo$, which proceeds via
a virtual top squark in the intermediate state.
Under the assumption of a 100\% branching fraction to this final state,
the cross section limit for each model point is compared with the theoretical
cross section to determine the excluded particle masses. Gluinos with a mass below 1600\GeV are excluded
at a 95$\%$ CL for scenarios with low $\PSGczDo$ mass, and neutralinos with a mass below 800\GeV are excluded for
a gluino mass of about 1300\GeV. In the T5tttt model, the top squark is lighter than the gluino, which
therefore decays via a two-body process. The boundary of the excluded region in the $\mGlu$-$\mLSP$ plane for T5tttt is found to
be only weakly sensitive to the top squark mass. These results significantly extend the sensitivity of single-lepton searches
based on data at $\sqrt{s} = 8\TeV$.

\begin{acknowledgments}

\hyphenation{Bundes-ministerium Forschungs-gemeinschaft Forschungs-zentren} We congratulate our colleagues in the CERN accelerator departments for the excellent performance of the LHC and thank the technical and administrative staffs at CERN and at other CMS institutes for their contributions to the success of the CMS effort. In addition, we gratefully acknowledge the computing centers and personnel of the Worldwide LHC Computing Grid for delivering so effectively the computing infrastructure essential to our analyses. Finally, we acknowledge the enduring support for the construction and operation of the LHC and the CMS detector provided by the following funding agencies: the Austrian Federal Ministry of Science, Research and Economy and the Austrian Science Fund; the Belgian Fonds de la Recherche Scientifique, and Fonds voor Wetenschappelijk Onderzoek; the Brazilian Funding Agencies (CNPq, CAPES, FAPERJ, and FAPESP); the Bulgarian Ministry of Education and Science; CERN; the Chinese Academy of Sciences, Ministry of Science and Technology, and National Natural Science Foundation of China; the Colombian Funding Agency (COLCIENCIAS); the Croatian Ministry of Science, Education and Sport, and the Croatian Science Foundation; the Research Promotion Foundation, Cyprus; the Ministry of Education and Research, Estonian Research Council via IUT23-4 and IUT23-6 and European Regional Development Fund, Estonia; the Academy of Finland, Finnish Ministry of Education and Culture, and Helsinki Institute of Physics; the Institut National de Physique Nucl\'eaire et de Physique des Particules~/~CNRS, and Commissariat \`a l'\'Energie Atomique et aux \'Energies Alternatives~/~CEA, France; the Bundesministerium f\"ur Bildung und Forschung, Deutsche Forschungsgemeinschaft, and Helmholtz-Gemeinschaft Deutscher Forschungszentren, Germany; the General Secretariat for Research and Technology, Greece; the National Scientific Research Foundation, and National Innovation Office, Hungary; the Department of Atomic Energy and the Department of Science and Technology, India; the Institute for Studies in Theoretical Physics and Mathematics, Iran; the Science Foundation, Ireland; the Istituto Nazionale di Fisica Nucleare, Italy; the Ministry of Science, ICT and Future Planning, and National Research Foundation (NRF), Republic of Korea; the Lithuanian Academy of Sciences; the Ministry of Education, and University of Malaya (Malaysia); the Mexican Funding Agencies (BUAP, CINVESTAV, CONACYT, LNS, SEP, and UASLP-FAI); the Ministry of Business, Innovation and Employment, New Zealand; the Pakistan Atomic Energy Commission; the Ministry of Science and Higher Education and the National Science Centre, Poland; the Funda\c{c}\~ao para a Ci\^encia e a Tecnologia, Portugal; JINR, Dubna; the Ministry of Education and Science of the Russian Federation, the Federal Agency of Atomic Energy of the Russian Federation, Russian Academy of Sciences, and the Russian Foundation for Basic Research; the Ministry of Education, Science and Technological Development of Serbia; the Secretar\'{\i}a de Estado de Investigaci\'on, Desarrollo e Innovaci\'on and Programa Consolider-Ingenio 2010, Spain; the Swiss Funding Agencies (ETH Board, ETH Zurich, PSI, SNF, UniZH, Canton Zurich, and SER); the Ministry of Science and Technology, Taipei; the Thailand Center of Excellence in Physics, the Institute for the Promotion of Teaching Science and Technology of Thailand, Special Task Force for Activating Research and the National Science and Technology Development Agency of Thailand; the Scientific and Technical Research Council of Turkey, and Turkish Atomic Energy Authority; the National Academy of Sciences of Ukraine, and State Fund for Fundamental Researches, Ukraine; the Science and Technology Facilities Council, UK; the US Department of Energy, and the US National Science Foundation.

Individuals have received support from the Marie-Curie program and the European Research Council and EPLANET (European Union); the Leventis Foundation; the A. P. Sloan Foundation; the Alexander von Humboldt Foundation; the Belgian Federal Science Policy Office; the Fonds pour la Formation \`a la Recherche dans l'Industrie et dans l'Agriculture (FRIA-Belgium); the Agentschap voor Innovatie door Wetenschap en Technologie (IWT-Belgium); the Ministry of Education, Youth and Sports (MEYS) of the Czech Republic; the Council of Science and Industrial Research, India; the HOMING PLUS program of the Foundation for Polish Science, cofinanced from European Union, Regional Development Fund; the Mobility Plus program of the Ministry of Science and Higher Education (Poland); the OPUS program of the National Science Center (Poland); the Thalis and Aristeia programs cofinanced by EU-ESF and the Greek NSRF; the National Priorities Research Program by Qatar National Research Fund; the Programa Clar\'in-COFUND del Principado de Asturias; the Rachadapisek Sompot Fund for Postdoctoral Fellowship, Chulalongkorn University (Thailand); the Chulalongkorn Academic into Its 2nd Century Project Advancement Project (Thailand); and the Welch Foundation, contract C-1845.

\end{acknowledgments}

\bibliography{auto_generated}

\providecommand{\href}[2]{#2}\begingroup\raggedright\begin{thebibliography}{10}%
\makeatletter
\providecommand{\hrefCMSnoop }[0]{\@secondoftwo}%
\makeatother
\providecommand{\doi}{\texttt{doi:}\begingroup \urlstyle{tt}\Url}

\bibitem{Ramond:1971gb}
\hrefCMSnoop {}{P.~Ramond, ``{Dual theory for free fermions}'',} \textit{ Phys.
  Rev. D} \textbf{ 3} (1971) 2415,
\href{http://dx.doi.org/10.1103/PhysRevD.3.2415}{\doi{10.1103/PhysRevD.3.2415}}.

\bibitem{Golfand:1971iw}
\href {http://www.jetpletters.ac.ru/ps/1584/article_24309.pdf}{Y.~A. Gol'fand
  and E.~P. Likhtman, ``{Extension of the algebra of Poincar\'{e} group
  generators and violation of P invariance}'',} \textit{ JETP Lett.} \textbf{
  13} (1971)
323.

\bibitem{Neveu:1971rx}
\hrefCMSnoop {}{A.~Neveu and J.~H. Schwarz, ``{Factorizable dual model of
  pions}'',} \textit{ Nucl. Phys. B} \textbf{ 31} (1971) 86,
\href{http://dx.doi.org/10.1016/0550-3213(71)90448-2}{\doi{10.1016/0550-3213(71)90448-2}}.

\bibitem{Volkov:1972jx}
\href {http://www.jetpletters.ac.ru/ps/1766/article_26864.pdf}{D.~V. Volkov and
  V.~P. Akulov, ``Possible universal neutrino interaction'',} \textit{ JETP
  Lett.} \textbf{ 16} (1972) 438.

\bibitem{Wess:1973kz}
\hrefCMSnoop {}{J.~Wess and B.~Zumino, ``{A {L}agrangian model invariant under
  supergauge transformations}'',} \textit{ Phys. Lett. B} \textbf{ 49} (1974)
  52,
\href{http://dx.doi.org/10.1016/0370-2693(74)90578-4}{\doi{10.1016/0370-2693(74)90578-4}}.

\bibitem{Wess:1974tw}
\hrefCMSnoop {}{J.~Wess and B.~Zumino, ``{Supergauge transformations in four
  dimensions}'',} \textit{ Nucl. Phys. B} \textbf{ 70} (1974) 39,
\href{http://dx.doi.org/10.1016/0550-3213(74)90355-1}{\doi{10.1016/0550-3213(74)90355-1}}.

\bibitem{Fayet:1974pd}
\hrefCMSnoop {}{P.~Fayet, ``{Supergauge invariant extension of the {H}iggs
  mechanism and a model for the electron and its neutrino}'',} \textit{ Nucl.
  Phys. B} \textbf{ 90} (1975) 104,
\href{http://dx.doi.org/10.1016/0550-3213(75)90636-7}{\doi{10.1016/0550-3213(75)90636-7}}.

\bibitem{Nilles:1983ge}
\hrefCMSnoop {}{H.~P. Nilles, ``{Supersymmetry, supergravity and particle
  physics}'',} \textit{ Phys. Rep.} \textbf{ 110} (1984) 1,
\href{http://dx.doi.org/10.1016/0370-1573(84)90008-5}{\doi{10.1016/0370-1573(84)90008-5}}.

\bibitem{tHooft:1979bh}
\href
  {https://inspirehep.net/record/144074/files/'t%20Hooft%20-%20Naturalness,%20Chiral%20Symmetry%20and%20Spontaneous%20Chiral%20Symmetry%20Breaking.pdf}{G.~'t~Hooft,
  ``Naturalness, chiral symmetry, and spontaneous chiral symmetry breaking'',}
  in \textit{ Recent Developments in Gauge Theories}, G.~'t~Hooft {et~al.},
  eds., p.~135.
\newblock Springer, 1980.
\newblock
NATO Advanced Study Institutes Series B.

\bibitem{Witten:1981nf}
\hrefCMSnoop {}{E.~Witten, ``{Dynamical breaking of supersymmetry}'',} \textit{
  Nucl. Phys. B} \textbf{ 188} (1981) 513,
\href{http://dx.doi.org/10.1016/0550-3213(81)90006-7}{\doi{10.1016/0550-3213(81)90006-7}}.

\bibitem{Dine:1981za}
\hrefCMSnoop {}{M.~Dine, W.~Fischler, and M.~Srednicki, ``{Supersymmetric
  technicolor}'',} \textit{ Nucl. Phys. B} \textbf{ 189} (1981) 575,
\href{http://dx.doi.org/10.1016/0550-3213(81)90582-4}{\doi{10.1016/0550-3213(81)90582-4}}.

\bibitem{Dimopoulos:1981au}
\hrefCMSnoop {}{S.~Dimopoulos and S.~Raby, ``{Supercolor}'',} \textit{ Nucl.
  Phys. B} \textbf{ 192} (1981) 353,
\href{http://dx.doi.org/10.1016/0550-3213(81)90430-2}{\doi{10.1016/0550-3213(81)90430-2}}.

\bibitem{Dimopoulos:1981zb}
\hrefCMSnoop {}{S.~Dimopoulos and H.~Georgi, ``{Softly broken supersymmetry and
  SU(5)}'',} \textit{ Nucl. Phys. B} \textbf{ 193} (1981) 150,
\href{http://dx.doi.org/10.1016/0550-3213(81)90522-8}{\doi{10.1016/0550-3213(81)90522-8}}.

\bibitem{Kaul:1981hi}
\hrefCMSnoop {}{R.~K. Kaul and P.~Majumdar, ``{Cancellation of quadratically
  divergent mass corrections in globally supersymmetric spontaneously broken
  gauge theories}'',} \textit{ Nucl. Phys. B} \textbf{ 199} (1982) 36,
\href{http://dx.doi.org/10.1016/0550-3213(82)90565-X}{\doi{10.1016/0550-3213(82)90565-X}}.

\bibitem{Zwicky:1933gu}
\hrefCMSnoop {}{F.~Zwicky, ``{Die Rotverschiebung von Extragalaktischen
  Nebeln}'',} \textit{ Helv. Phys. Acta} \textbf{ 6} (1933)
110.

\bibitem{Rubin:1970zza}
\hrefCMSnoop {}{V.~C. Rubin and W.~K. Ford~Jr, ``{Rotation of the Andromeda
  nebula from a spectroscopic survey of emission regions}'',} \textit{
  Astrophys. J.} \textbf{ 159} (1970) 379,
\href{http://dx.doi.org/10.1086/150317}{\doi{10.1086/150317}}.

\bibitem{Agashe:2014kda}
\hrefCMSnoop {}{{Particle Data Group}, K.~A. Olive {et~al.}, ``{Review of
  Particle Physics}'',} \textit{ Chin. Phys. C} \textbf{ 38} (2014) 090001,
\href{http://dx.doi.org/10.1088/1674-1137/38/9/090001}{\doi{10.1088/1674-1137/38/9/090001}}.

\bibitem{PhysRevD.24.1681}
\hrefCMSnoop {}{S.~Dimopoulos, S.~Raby, and F.~Wilczek, ``Supersymmetry and the
  scale of unification'',} \textit{ Phys. Rev. D} \textbf{ 24} (1981) 1681,
  \href{http://dx.doi.org/10.1103/PhysRevD.24.1681}{\doi{10.1103/PhysRevD.24.1681}}.

\bibitem{Sakai1981}
\hrefCMSnoop {}{N.~Sakai, ``Naturalness in supersymmetric GUTS'',} \textit{ Z.
  Phys. C} \textbf{ 11} (1981) 153,
  \href{http://dx.doi.org/10.1007/BF01573998}{\doi{10.1007/BF01573998}}.

\bibitem{IBANEZ1981439}
\hrefCMSnoop {}{L.~E. Ib{\'a}{\~n}ez and G.~G. Ross, ``Low-energy predictions
  in supersymmetric grand unified theories'',} \textit{ Phys. Lett. B} \textbf{
  105} (1981) 439,
  \href{http://dx.doi.org/10.1016/0370-2693(81)91200-4}{\doi{10.1016/0370-2693(81)91200-4}}.

\bibitem{EINHORN1982475}
\hrefCMSnoop {}{M.~B. Einhorn and D.~R.~T. Jones, ``The weak mixing angle and
  unification mass in supersymmetric SU(5)'',} \textit{ Nucl. Phys. B} \textbf{
  196} (1982) 475,
  \href{http://dx.doi.org/10.1016/0550-3213(82)90502-8}{\doi{10.1016/0550-3213(82)90502-8}}.

\bibitem{PhysRevD.25.3092}
\hrefCMSnoop {}{W.~J. Marciano and G.~Senjanovi{\'c}, ``Predictions of
  supersymmetric grand unified theories'',} \textit{ Phys. Rev. D} \textbf{ 25}
  (Jun, 1982) 3092,
  \href{http://dx.doi.org/10.1103/PhysRevD.25.3092}{\doi{10.1103/PhysRevD.25.3092}}.

\bibitem{Farrar:1978xj}
\hrefCMSnoop {}{G.~R. Farrar and P.~Fayet, ``{Phenomenology of the production,
  decay, and detection of new hadronic states associated with
  supersymmetry}'',} \textit{ Phys. Lett. B} \textbf{ 76} (1978) 575,
\href{http://dx.doi.org/10.1016/0370-2693(78)90858-4}{\doi{10.1016/0370-2693(78)90858-4}}.

\bibitem{Martin:1997ns}
\hrefCMSnoop {}{S.~P. Martin, ``A Supersymmetry Primer'',} in \textit{
  Perspectives on Supersymmetry {II}}, G.~L. Kane, ed., p.~1.
\newblock 2010.
\newblock
  \href{http://www.arXiv.org/abs/hep-ph/9709356}{\texttt{arXiv:hep-ph/9709356}}.
\newblock Adv. Ser. Direct. High Energy Phys., vol. 21.
\href{http://dx.doi.org/10.1142/9789814307505_0001}{\doi{10.1142/9789814307505_0001}}.

\bibitem{Aad:2012tfa}
\hrefCMSnoop {}{{ATLAS Collaboration}, ``{Observation of a new particle in the
  search for the Standard Model Higgs boson with the ATLAS detector at the
  LHC}'',} \textit{ Phys. Lett. B} \textbf{ 716} (2012) 1,
  \href{http://dx.doi.org/10.1016/j.physletb.2012.08.020}{\doi{10.1016/j.physletb.2012.08.020}},
\href{http://www.arXiv.org/abs/1207.7214}{\texttt{arXiv:1207.7214}}.

\bibitem{Chatrchyan:2012ufa}
\hrefCMSnoop {}{{CMS Collaboration}, ``{Observation of a new boson at a mass of
  125 GeV with the CMS experiment at the LHC}'',} \textit{ Phys. Lett. B}
  \textbf{ 716} (2012) 30,
  \href{http://dx.doi.org/10.1016/j.physletb.2012.08.021}{\doi{10.1016/j.physletb.2012.08.021}},
\href{http://www.arXiv.org/abs/1207.7235}{\texttt{arXiv:1207.7235}}.

\bibitem{Chatrchyan:2013lba}
\hrefCMSnoop {}{{CMS Collaboration}, ``{Observation of a new boson with mass
  near 125\GeV in pp collisions at $\sqrt{s}$ = 7 and 8\TeV}'',} \textit{ JHEP}
  \textbf{ 06} (2013) 081,
  \href{http://dx.doi.org/10.1007/JHEP06(2013)081}{\doi{10.1007/JHEP06(2013)081}},
\href{http://www.arXiv.org/abs/1303.4571}{\texttt{arXiv:1303.4571}}.

\bibitem{Khachatryan:2014jba}
\hrefCMSnoop {}{{CMS Collaboration}, ``{Precise determination of the mass of
  the Higgs boson and tests of compatibility of its couplings with the standard
  model predictions using proton collisions at 7 and 8\TeV }'',} \textit{ Eur.
  Phys. J. C} \textbf{ 75} (2015) 212,
  \href{http://dx.doi.org/10.1140/epjc/s10052-015-3351-7}{\doi{10.1140/epjc/s10052-015-3351-7}},
\href{http://www.arXiv.org/abs/1412.8662}{\texttt{arXiv:1412.8662}}.

\bibitem{Aad:2014aba}
\hrefCMSnoop {}{{ATLAS Collaboration}, ``{Measurement of the Higgs boson mass
  from the $\textrm{H}\rightarrow \gamma\gamma$ and $\textrm{H} \rightarrow
  \textrm{ZZ}^{*} \rightarrow 4\ell$ channels with the ATLAS detector using 25
  fb$^{-1}$ of pp collision data}'',} \textit{ Phys. Rev. D} \textbf{ 90}
  (2014) 052004,
  \href{http://dx.doi.org/10.1103/PhysRevD.90.052004}{\doi{10.1103/PhysRevD.90.052004}},
\href{http://www.arXiv.org/abs/1406.3827}{\texttt{arXiv:1406.3827}}.

\bibitem{Aad:2015zhl}
\hrefCMSnoop {}{{{ATLAS and CMS Collaborations}}, ``{Combined measurement of
  the Higgs boson mass in pp collisions at $\sqrt{s} = 7$ and 8\TeV with the
  ATLAS and CMS experiments}'',} \textit{ Phys. Rev. Lett.} \textbf{ 114}
  (2015) 191803,
  \href{http://dx.doi.org/10.1103/PhysRevLett.114.191803}{\doi{10.1103/PhysRevLett.114.191803}},
\href{http://www.arXiv.org/abs/1503.07589}{\texttt{arXiv:1503.07589}}.

\bibitem{Dimopoulos:1995mi}
\hrefCMSnoop {}{S.~Dimopoulos and G.~F. Giudice, ``{Naturalness constraints in
  supersymmetric theories with nonuniversal soft terms}'',} \textit{ Phys.
  Lett. B} \textbf{ 357} (1995) 573,
  \href{http://dx.doi.org/10.1016/0370-2693(95)00961-J}{\doi{10.1016/0370-2693(95)00961-J}},
\href{http://www.arXiv.org/abs/hep-ph/9507282}{\texttt{arXiv:hep-ph/9507282}}.

\bibitem{Barbieri:2009ev}
\hrefCMSnoop {}{R.~Barbieri and D.~Pappadopulo, ``{S-particles at their
  naturalness limits}'',} \textit{ JHEP} \textbf{ 10} (2009) 061,
  \href{http://dx.doi.org/10.1088/1126-6708/2009/10/061}{\doi{10.1088/1126-6708/2009/10/061}},
\href{http://www.arXiv.org/abs/0906.4546}{\texttt{arXiv:0906.4546}}.

\bibitem{Papucci:2011wy}
\hrefCMSnoop {}{M.~Papucci, J.~T. Ruderman, and A.~Weiler, ``{Natural {SUSY}
  endures}'',} \textit{ JHEP} \textbf{ 09} (2012) 035,
  \href{http://dx.doi.org/10.1007/JHEP09(2012)035}{\doi{10.1007/JHEP09(2012)035}},
\href{http://www.arXiv.org/abs/1110.6926}{\texttt{arXiv:1110.6926}}.

\bibitem{Feng:2013pwa}
\hrefCMSnoop {}{J.~L. Feng, ``{Naturalness and the status of supersymmetry}'',}
  \textit{ Ann. Rev. Nucl. Part. Sci.} \textbf{ 63} (2013) 351,
  \href{http://dx.doi.org/10.1146/annurev-nucl-102010-130447}{\doi{10.1146/annurev-nucl-102010-130447}},
\href{http://www.arXiv.org/abs/1302.6587}{\texttt{arXiv:1302.6587}}.

\bibitem{Craig:2013cxa}
\href
  {http://inspirehep.net/record/1252552/files/arXiv:1309.0528.pdf}{N.~Craig,
  ``{The state of supersymmetry after Run I of the LHC}'',} in \textit{ {Beyond
  the Standard Model after the first run of the LHC, Arcetri, Florence, Italy,
  May 20-July 12, 2013}}.
\newblock 2013.
\newblock
\href{http://www.arXiv.org/abs/1309.0528}{\texttt{arXiv:1309.0528}}.
\newblock

\bibitem{Aad:2014lra}
\hrefCMSnoop {}{{ATLAS Collaboration}, ``{Search for strong production of
  supersymmetric particles in final states with missing transverse momentum and
  at least three $b$-jets at $\sqrt{s}$= 8 TeV proton-proton collisions with
  the ATLAS detector}'',} \textit{ JHEP} \textbf{ 10} (2014) 024,
  \href{http://dx.doi.org/10.1007/JHEP10(2014)024}{\doi{10.1007/JHEP10(2014)024}},
\href{http://www.arXiv.org/abs/1407.0600}{\texttt{arXiv:1407.0600}}.

\bibitem{Aad:2015mia}
\hrefCMSnoop {}{{ATLAS Collaboration}, ``{Search for squarks and gluinos in
  events with isolated leptons, jets and missing transverse momentum at
  $\sqrt{s}=8\TeV$ with the ATLAS detector}'',} \textit{ JHEP} \textbf{ 04}
  (2015) 116,
  \href{http://dx.doi.org/10.1007/JHEP04(2015)116}{\doi{10.1007/JHEP04(2015)116}},
\href{http://www.arXiv.org/abs/1501.03555}{\texttt{arXiv:1501.03555}}.

\bibitem{Chatrchyan:2013iqa}
\hrefCMSnoop {}{{CMS Collaboration}, ``{Search for supersymmetry in pp
  collisions at $\sqrt{s} = 8\TeV$ in events with a single lepton, large jet
  multiplicity, and multiple b jets}'',} \textit{ Phys. Lett. B} \textbf{ 733}
  (2014) 328,
  \href{http://dx.doi.org/10.1016/j.physletb.2014.04.023}{\doi{10.1016/j.physletb.2014.04.023}},
\href{http://www.arXiv.org/abs/1311.4937}{\texttt{arXiv:1311.4937}}.

\bibitem{Borschensky:2014cia}
C.~Borschensky\hrefCMSnoop {}{ {et~al.}, ``{Squark and gluino production cross
  sections in pp collisions at $\sqrt{s} = 13$, 14, 33 and 100\TeV}'',}
  \textit{ Eur. Phys. J. C} \textbf{ 74} (2014) 3174,
  \href{http://dx.doi.org/10.1140/epjc/s10052-014-3174-y}{\doi{10.1140/epjc/s10052-014-3174-y}},
\href{http://www.arXiv.org/abs/1407.5066}{\texttt{arXiv:1407.5066}}.

\bibitem{ref:ttbarXSec}
\hrefCMSnoop {}{M.~Czakon, P.~Fiedler, and A.~Mitov, ``{The total top quark
  pair production cross-section at $\mathcal{O}(\alpha^4_S)$}'',} \textit{
  Phys. Rev. Lett.} \textbf{ 110} (2013) 252004,
  \href{http://dx.doi.org/10.1103/PhysRevLett.110.252004}{\doi{10.1103/PhysRevLett.110.252004}},
  \href{http://www.arXiv.org/abs/1303.6254}{\texttt{arXiv:1303.6254}}.

\bibitem{Chatrchyan:2013sza}
\hrefCMSnoop {}{{CMS Collaboration}, ``{Interpretation of searches for
  supersymmetry with simplified models}'',} \textit{ Phys. Rev. D} \textbf{ 88}
  (2013) 052017,
  \href{http://dx.doi.org/10.1103/PhysRevD.88.052017}{\doi{10.1103/PhysRevD.88.052017}},
\href{http://www.arXiv.org/abs/1301.2175}{\texttt{arXiv:1301.2175}}.

\bibitem{bib-sms-2}
\hrefCMSnoop {}{J.~Alwall, P.~Schuster, and N.~Toro, ``Simplified models for a
  first characterization of new physics at the {LHC}'',} \textit{ Phys. Rev. D}
  \textbf{ 79} (2009) 075020,
  \href{http://dx.doi.org/10.1103/PhysRevD.79.075020}{\doi{10.1103/PhysRevD.79.075020}},
\href{http://www.arXiv.org/abs/0810.3921}{\texttt{arXiv:0810.3921}}.

\bibitem{bib-sms-3}
\hrefCMSnoop {}{J.~Alwall, M.-P. Le, M.~Lisanti, and J.~G. Wacker,
  ``{Model-independent jets plus missing energy searches}'',} \textit{ Phys.
  Rev. D} \textbf{ 79} (2009) 015005,
  \href{http://dx.doi.org/10.1103/PhysRevD.79.015005}{\doi{10.1103/PhysRevD.79.015005}},
\href{http://www.arXiv.org/abs/0809.3264}{\texttt{arXiv:0809.3264}}.

\bibitem{bib-sms-4}
D.~Alves\hrefCMSnoop {}{ {et~al.}, ``Simplified models for {LHC} new physics
  searches'',} \textit{ J. Phys. G} \textbf{ 39} (2012) 105005,
  \href{http://dx.doi.org/10.1088/0954-3899/39/10/105005}{\doi{10.1088/0954-3899/39/10/105005}},
\href{http://www.arXiv.org/abs/1105.2838}{\texttt{arXiv:1105.2838}}.

\bibitem{Hook:2012fd}
\hrefCMSnoop {}{A.~Hook, E.~Izaguirre, M.~Lisanti, and J.~G. Wacker, ``{High
  multiplicity searches at the LHC using jet masses}'',} \textit{ Phys. Rev. D}
  \textbf{ 85} (2012) 055029,
  \href{http://dx.doi.org/10.1103/PhysRevD.85.055029}{\doi{10.1103/PhysRevD.85.055029}},
\href{http://www.arXiv.org/abs/1202.0558}{\texttt{arXiv:1202.0558}}.

\bibitem{Cohen:2012yc}
\hrefCMSnoop {}{T.~Cohen, E.~Izaguirre, M.~Lisanti, and H.~K. Lou, ``{Jet
  substructure by accident}'',} \textit{ JHEP} \textbf{ 03} (2013) 161,
  \href{http://dx.doi.org/10.1007/JHEP03(2013)161}{\doi{10.1007/JHEP03(2013)161}},
\href{http://www.arXiv.org/abs/1212.1456}{\texttt{arXiv:1212.1456}}.

\bibitem{Hedri:2013pvl}
\hrefCMSnoop {}{S.~El~Hedri, A.~Hook, M.~Jankowiak, and J.~G. Wacker,
  ``{Learning how to count: a high multiplicity search for the LHC}'',}
  \textit{ JHEP} \textbf{ 08} (2013) 136,
  \href{http://dx.doi.org/10.1007/JHEP08(2013)136}{\doi{10.1007/JHEP08(2013)136}},
\href{http://www.arXiv.org/abs/1302.1870}{\texttt{arXiv:1302.1870}}.

\bibitem{Aad:2015lea}
\hrefCMSnoop {}{{ATLAS Collaboration}, ``{Search for massive supersymmetric
  particles decaying to many jets using the ATLAS detector in pp collisions at
  $\sqrt{s} = 8\TeV$}'',} \textit{ Phys. Rev. D} \textbf{ 91} (2015) 112016,
  \href{http://dx.doi.org/10.1103/PhysRevD.91.112016}{\doi{10.1103/PhysRevD.91.112016}},
\href{http://www.arXiv.org/abs/1502.05686}{\texttt{arXiv:1502.05686}}.

\bibitem{Aad:2013wta}
\hrefCMSnoop {}{{ATLAS Collaboration}, ``{Search for new phenomena in final
  states with large jet multiplicities and missing transverse momentum at
  $\sqrt{s}=8\TeV$ proton-proton collisions using the ATLAS experiment}'',}
  \textit{ JHEP} \textbf{ 10} (2013) 130,
  \href{http://dx.doi.org/10.1007/JHEP10(2013)130}{\doi{10.1007/JHEP10(2013)130}},
  \href{http://www.arXiv.org/abs/1308.1841}{\texttt{arXiv:1308.1841}}.
[Erratum: \DOI{10.1007/JHEP01(2014)109}].

\bibitem{SUSYDPS}
\href {https://cds.cern.ch/record/2049757}{{CMS Collaboration},
  ``{Commissioning the performance of key observables used in SUSY searches
  with the first 13\TeV data}'',} CMS Detector Performance Report
  CMS-DP-2015-035, CERN, 2015.

\bibitem{Chatrchyan:2008zzk}
\hrefCMSnoop {}{{CMS Collaboration}, ``The {CMS} experiment at the {CERN}
  {LHC}'',} \textit{ JINST} \textbf{ 3} (2008) S08004,
\href{http://dx.doi.org/10.1088/1748-0221/3/08/S08004}{\doi{10.1088/1748-0221/3/08/S08004}}.

\bibitem{Alwall:2014hca}
J.~Alwall\hrefCMSnoop {}{ {et~al.}, ``{The automated computation of tree-level
  and next-to-leading order differential cross sections, and their matching to
  parton shower simulations}'',} \textit{ JHEP} \textbf{ 07} (2014) 079,
  \href{http://dx.doi.org/10.1007/JHEP07(2014)079}{\doi{10.1007/JHEP07(2014)079}},
\href{http://www.arXiv.org/abs/1405.0301}{\texttt{arXiv:1405.0301}}.

\bibitem{powheg-singletop-tchan}
\hrefCMSnoop {}{S.~Alioli, P.~Nason, C.~Oleari, and E.~Re, ``{NLO single-top
  production matched with shower in POWHEG: $s$- and $t$-channel
  contributions}'',} \textit{ JHEP} \textbf{ 09} (2009) 111,
  \href{http://dx.doi.org/10.1088/1126-6708/2009/09/111}{\doi{10.1088/1126-6708/2009/09/111}},
  \href{http://www.arXiv.org/abs/0907.4076}{\texttt{arXiv:0907.4076}}.
[Erratum: \DOI{10.1007/JHEP02(2010)011}].

\bibitem{powheg-singletop-wt}
\hrefCMSnoop {}{E.~Re, ``{Single-top $\PW\PQt$-channel production matched with
  parton showers using the POWHEG method}'',} \textit{ Eur. Phys. J. C}
  \textbf{ 71} (2011) 1547,
  \href{http://dx.doi.org/10.1140/epjc/s10052-011-1547-z}{\doi{10.1140/epjc/s10052-011-1547-z}},
\href{http://www.arXiv.org/abs/1009.2450}{\texttt{arXiv:1009.2450}}.

\bibitem{Ball:2014uwa}
\hrefCMSnoop {}{{NNPDF} Collaboration, ``{Parton distributions for the LHC Run
  II}'',} \textit{ JHEP} \textbf{ 04} (2015) 040,
  \href{http://dx.doi.org/10.1007/JHEP04(2015)040}{\doi{10.1007/JHEP04(2015)040}},
\href{http://www.arXiv.org/abs/1410.8849}{\texttt{arXiv:1410.8849}}.

\bibitem{Sjostrand:2014zea}
T.~Sj{\"o}strand\hrefCMSnoop {}{ {et~al.}, ``{An Introduction to PYTHIA
  8.2}'',} \textit{ Comput. Phys. Commun.} \textbf{ 191} (2015) 159,
  \href{http://dx.doi.org/10.1016/j.cpc.2015.01.024}{\doi{10.1016/j.cpc.2015.01.024}},
\href{http://www.arXiv.org/abs/1410.3012}{\texttt{arXiv:1410.3012}}.

\bibitem{Khachatryan:2110213}
\hrefCMSnoop {}{{CMS Collaboration}, ``{Event generator tunes obtained from
  underlying event and multiparton scattering measurements}'',} \textit{ Eur.
  Phys. J. C} \textbf{ 76} (2016), no.~3, 155,
  \href{http://dx.doi.org/10.1140/epjc/s10052-016-3988-x}{\doi{10.1140/epjc/s10052-016-3988-x}},
\href{http://www.arXiv.org/abs/1512.00815}{\texttt{arXiv:1512.00815}}.

\bibitem{Agostinelli:2002hh}
\hrefCMSnoop {}{{GEANT4} Collaboration, ``{GEANT4}---a simulation toolkit'',}
  \textit{ Nucl. Instrum. Meth. A} \textbf{ 506} (2003) 250,
\href{http://dx.doi.org/10.1016/S0168-9002(03)01368-8}{\doi{10.1016/S0168-9002(03)01368-8}}.

\bibitem{Khachatryan:2016641}
\hrefCMSnoop {}{{CMS Collaboration}, ``{Measurement of the differential cross
  section for top quark pair production in pp collisions at $\sqrt{s} =
  8\TeV$}'',} \textit{ Eur. Phys. J. C} \textbf{ 75} (2015) 542,
  \href{http://dx.doi.org/10.1140/epjc/s10052-015-3709-x}{\doi{10.1140/epjc/s10052-015-3709-x}},
\href{http://www.arXiv.org/abs/1505.04480}{\texttt{arXiv:1505.04480}}.

\bibitem{Abdullin:2011zz}
\hrefCMSnoop {}{{CMS Collaboration}, ``{The fast simulation of the CMS detector
  at LHC}'',} \textit{ J. Phys. Conf. Ser.} \textbf{ 331} (2011) 032049,
\href{http://dx.doi.org/10.1088/1742-6596/331/3/032049}{\doi{10.1088/1742-6596/331/3/032049}}.

\bibitem{Beenakker:1996ch}
\hrefCMSnoop {}{W.~Beenakker, R.~Hopker, M.~Spira, and P.~M. Zerwas, ``{Squark
  and gluino production at hadron colliders}'',} \textit{ Nucl. Phys. B}
  \textbf{ 492} (1997) 51,
  \href{http://dx.doi.org/10.1016/S0550-3213(97)00084-9}{\doi{10.1016/S0550-3213(97)00084-9}},
\href{http://www.arXiv.org/abs/hep-ph/9610490}{\texttt{arXiv:hep-ph/9610490}}.

\bibitem{Kulesza:2008jb}
\hrefCMSnoop {}{A.~Kulesza and L.~Motyka, ``{Threshold resummation for
  squark-antisquark and gluino-pair production at the LHC}'',} \textit{ Phys.
  Rev. Lett.} \textbf{ 102} (2009) 111802,
  \href{http://dx.doi.org/10.1103/PhysRevLett.102.111802}{\doi{10.1103/PhysRevLett.102.111802}},
\href{http://www.arXiv.org/abs/0807.2405}{\texttt{arXiv:0807.2405}}.

\bibitem{Kulesza:2009kq}
\hrefCMSnoop {}{A.~Kulesza and L.~Motyka, ``{Soft gluon resummation for the
  production of gluino-gluino and squark-antisquark pairs at the LHC}'',}
  \textit{ Phys. Rev. D} \textbf{ 80} (2009) 095004,
  \href{http://dx.doi.org/10.1103/PhysRevD.80.095004}{\doi{10.1103/PhysRevD.80.095004}},
\href{http://www.arXiv.org/abs/0905.4749}{\texttt{arXiv:0905.4749}}.

\bibitem{Beenakker:2009ha}
W.~Beenakker\hrefCMSnoop {}{ {et~al.}, ``Soft-gluon resummation for squark and
  gluino hadroproduction'',} \textit{ JHEP} \textbf{ 12} (2009) 041,
  \href{http://dx.doi.org/10.1088/1126-6708/2009/12/041}{\doi{10.1088/1126-6708/2009/12/041}},
\href{http://www.arXiv.org/abs/0909.4418}{\texttt{arXiv:0909.4418}}.

\bibitem{cms-pas-pft-09-001}
\href {http://cdsweb.cern.ch/record/1194487}{{CMS Collaboration}, ``Particle
  flow event reconstruction in {CMS} and performance for jets, taus and
  \MET'',} CMS Physics Analysis Summary CMS-PAS-PFT-09-001, CERN, 2009.

\bibitem{cms-pas-pft-10-001}
\href {http://cdsweb.cern.ch/record/1247373}{{CMS Collaboration},
  ``Commissioning of the particle-flow event reconstruction with the first LHC
  collisions recorded in the CMS detector'',} CMS Physics Analysis Summary
  CMS-PAS-PFT-10-001, CERN, 2010.

\bibitem{Cacciari:2008gp}
\hrefCMSnoop {}{M.~Cacciari, G.~P. Salam, and G.~Soyez, ``The anti-$k_{\rm t}$
  jet clustering algorithm'',} \textit{ JHEP} \textbf{ 04} (2008) 063,
  \href{http://dx.doi.org/10.1088/1126-6708/2008/04/063}{\doi{10.1088/1126-6708/2008/04/063}},
  \href{http://www.arXiv.org/abs/0802.1189}{\texttt{arXiv:0802.1189}}.

\bibitem{Cacciari:2011ma}
\hrefCMSnoop {}{M.~Cacciari, G.~P. Salam, and G.~Soyez, ``{FastJet user
  manual}'',} \textit{ Eur. Phys. J. C} \textbf{ 72} (2012) 1896,
  \href{http://dx.doi.org/10.1140/epjc/s10052-012-1896-2}{\doi{10.1140/epjc/s10052-012-1896-2}},
\href{http://www.arXiv.org/abs/1111.6097}{\texttt{arXiv:1111.6097}}.

\bibitem{Cacciari:2007fd}
\hrefCMSnoop {}{M.~Cacciari and G.~P. Salam, ``{Pileup subtraction using jet
  areas}'',} \textit{ Phys. Lett. B} \textbf{ 659} (2008) 119,
  \href{http://dx.doi.org/10.1016/j.physletb.2007.09.077}{\doi{10.1016/j.physletb.2007.09.077}},
\href{http://www.arXiv.org/abs/0707.1378}{\texttt{arXiv:0707.1378}}.

\bibitem{Chatrchyan:2011ds}
\hrefCMSnoop {}{{CMS Collaboration}, ``{Determination of jet energy calibration
  and transverse momentum resolution in {CMS}}'',} \textit{ JINST} \textbf{ 6}
  (2011) P11002,
  \href{http://dx.doi.org/10.1088/1748-0221/6/11/P11002}{\doi{10.1088/1748-0221/6/11/P11002}},
\href{http://www.arXiv.org/abs/1107.4277}{\texttt{arXiv:1107.4277}}.

\bibitem{Chatrchyan:2012jua}
\hrefCMSnoop {}{{CMS Collaboration}, ``{Identification of $\PQb$-quark jets
  with the CMS experiment}'',} \textit{ JINST} \textbf{ 8} (2013) P04013,
  \href{http://dx.doi.org/10.1088/1748-0221/8/04/P04013}{\doi{10.1088/1748-0221/8/04/P04013}},
\href{http://www.arXiv.org/abs/1211.4462}{\texttt{arXiv:1211.4462}}.

\bibitem{CMS:2016kkf}
\href {https://cds.cern.ch/record/2138504}{{CMS Collaboration},
  ``{Identification of $\PQb$ quark jets at the CMS Experiment in the LHC Run
  2}'',} CMS Physics Analysis Summary CMS-PAS-BTV-15-001, CERN, 2016.

\bibitem{Khachatryan:2015hwa}
\hrefCMSnoop {}{{CMS Collaboration}, ``{Performance of electron reconstruction
  and selection with the CMS detector in proton-proton collisions at $\sqrt{s}
  = 8\TeV$}'',} \textit{ JINST} \textbf{ 10} (2015) P06005,
  \href{http://dx.doi.org/10.1088/1748-0221/10/06/P06005}{\doi{10.1088/1748-0221/10/06/P06005}},
\href{http://www.arXiv.org/abs/1502.02701}{\texttt{arXiv:1502.02701}}.

\bibitem{Chatrchyan:2012xi}
\hrefCMSnoop {}{{CMS Collaboration}, ``{Performance of CMS muon reconstruction
  in pp collision events at $\sqrt{s}=7\TeV$}'',} \textit{ JINST} \textbf{ 7}
  (2012) P10002,
  \href{http://dx.doi.org/10.1088/1748-0221/7/10/P10002}{\doi{10.1088/1748-0221/7/10/P10002}},
\href{http://www.arXiv.org/abs/1206.4071}{\texttt{arXiv:1206.4071}}.

\bibitem{Rehermann:2010vq}
\hrefCMSnoop {}{K.~Rehermann and B.~Tweedie, ``{Efficient identification of
  boosted semileptonic top quarks at the LHC}'',} \textit{ JHEP} \textbf{ 03}
  (2011) 059,
  \href{http://dx.doi.org/10.1007/JHEP03(2011)059}{\doi{10.1007/JHEP03(2011)059}},
\href{http://www.arXiv.org/abs/1007.2221}{\texttt{arXiv:1007.2221}}.

\bibitem{Aad:2014bva}
\hrefCMSnoop {}{{ATLAS Collaboration}, ``{Search for direct pair production of
  the top squark in all-hadronic final states in proton-proton collisions at
  $\sqrt{s}=8\TeV$ with the ATLAS detector}'',} \textit{ JHEP} \textbf{ 09}
  (2014) 015,
  \href{http://dx.doi.org/10.1007/JHEP09(2014)015}{\doi{10.1007/JHEP09(2014)015}},
\href{http://www.arXiv.org/abs/1406.1122}{\texttt{arXiv:1406.1122}}.

\bibitem{Khachatryan:2133129}
\href {http://cds.cern.ch/record/2133129}{{CMS Collaboration}, ``{Search for
  supersymmetry in the multijet and missing transverse momentum final state in
  pp collisions at 13\TeV }'',} (2016).
  \href{http://www.arXiv.org/abs/1602.06581}{\texttt{arXiv:1602.06581}}.
  Submitted to Phys. Lett. B.

\bibitem{CMS-PAS-LUM-15-001}
\href {https://cds.cern.ch/record/2138682}{{CMS Collaboration}, ``{CMS
  luminosity measurement for the 2015 data taking period}'',} CMS Physics
  Analysis Summary CMS-PAS-LUM-15-001, CERN, Geneva, 2016.

\bibitem{Junk:1999kv}
\hrefCMSnoop {}{T.~Junk, ``{Confidence level computation for combining searches
  with small statistics}'',} \textit{ Nucl. Instrum. Meth. A} \textbf{ 434}
  (1999) 435,
  \href{http://dx.doi.org/10.1016/S0168-9002(99)00498-2}{\doi{10.1016/S0168-9002(99)00498-2}},
\href{http://www.arXiv.org/abs/hep-ex/9902006}{\texttt{arXiv:hep-ex/9902006}}.

\bibitem{0954-3899-28-10-313}
\hrefCMSnoop {}{A.~L. Read, ``{Presentation of search results: the {$\rm CL_s$}
  technique}'',} \textit{ J. Phys. G} \textbf{ 28} (2002) 2693,
\href{http://dx.doi.org/10.1088/0954-3899/28/10/313}{\doi{10.1088/0954-3899/28/10/313}}.

\bibitem{CMS-NOTE-2011-005}
\href {https://cds.cern.ch/record/1379837}{{ATLAS Collaboration, CMS
  Collaboration, LHC Higgs Combination Group}, ``{Procedure for the LHC Higgs
  boson search combination in Summer 2011}'',} Technical Report
  CMS-NOTE-2011-005, ATL-PHYS-PUB-2011-11, CERN, 2011.

\end{thebibliography}\endgroup

\cleardoublepage \appendix\section{The CMS Collaboration \label{app:collab}}\begin{sloppypar}\hyphenpenalty=5000\widowpenalty=500\clubpenalty=5000\textbf{Yerevan Physics Institute,  Yerevan,  Armenia}\\*[0pt]
V.~Khachatryan, A.M.~Sirunyan, A.~Tumasyan
\vskip\cmsinstskip
\textbf{Institut f\"{u}r Hochenergiephysik der OeAW,  Wien,  Austria}\\*[0pt]
W.~Adam, E.~Asilar, T.~Bergauer, J.~Brandstetter, E.~Brondolin, M.~Dragicevic, J.~Er\"{o}, M.~Flechl, M.~Friedl, R.~Fr\"{u}hwirth\cmsAuthorMark{1}, V.M.~Ghete, C.~Hartl, N.~H\"{o}rmann, J.~Hrubec, M.~Jeitler\cmsAuthorMark{1}, A.~K\"{o}nig, I.~Kr\"{a}tschmer, D.~Liko, T.~Matsushita, I.~Mikulec, D.~Rabady, N.~Rad, B.~Rahbaran, H.~Rohringer, J.~Schieck\cmsAuthorMark{1}, J.~Strauss, W.~Treberer-Treberspurg, W.~Waltenberger, C.-E.~Wulz\cmsAuthorMark{1}
% ...
\vskip\cmsinstskip
% ...
1:~~Also at Vienna University of Technology, Vienna, Austria\\
% ...

\end{sloppypar}
\end{document}